\documentclass[useAMS,usenatbib]{mn2e}
\usepackage{hyperref}
\usepackage{graphicx}
\usepackage{lscape}
\usepackage{url} 
\usepackage[dvips]{color}
\usepackage{array}
\usepackage{aas_macros}
\usepackage{graphicx}
\usepackage{mathtools}
\DeclarePairedDelimiter\abs{\lvert}{\rvert}

\usepackage{times}
\usepackage{amsmath}
\usepackage[raggedright,FIGTOPCAP]{subfigure}
\usepackage{lineno}
\usepackage[T1]{fontenc}
\usepackage{aecompl} 
\newcolumntype{x}[1]{>{\centering\let\newline\\\arraybackslash\hspace{0pt}}p{#1}}
\newcolumntype{R}[1]{>{\raggedleft\let\newline\\\arraybackslash\hspace{0pt}}p{#1}}
\newcolumntype{L}[1]{>{\raggedright\let\newline\\\arraybackslash\hspace{0pt}}p{#1}}
\usepackage{ragged2e}
\newcolumntype{P}[1]{>{\RaggedRight\hspace{0pt}}p{#1}}

\newcommand{\fermimin}{2008 August 4}
\newcommand{\fermimax}{2013 May 1}

\title[HTRU X: Timing solutions for 16 MSPs]
  {The High Time Resolution Universe Pulsar Survey X: Discovery of four millisecond pulsars and updated timing solutions of a further 12}
\author[C. Ng et al.]
  {C.~Ng$^1$, M.~Bailes$^{2,3}$, S.~D.~Bates$^4$, N.~D.~R.~Bhat$^{2,3,5}$, M.~Burgay$^6$, S.~Burke-Spolaor$^7$, 
  \newauthor
  D.~J.~Champion$^1$, P.~Coster$^{2,8}$, S.~Johnston$^8$, M.~J.~Keith$^8$, M.~Kramer$^{1,9}$, L.~Levin$^4$,
  \newauthor
  E.~Petroff$^{2,3,8}$, A.~Possenti$^6$, B.~W.~Stappers$^9$, W.~van~Straten$^{2,3}$, D.~Thornton$^{8,9}$, 
  \newauthor
  C.~Tiburzi$^{6,10}$, C.~G.~Bassa$^9$, P.~C.~C.~Freire$^1$, L.~Guillemot$^{1,11}$, A.~G.~Lyne$^9$,
  \newauthor
  T.~M.~Tauris$^{12,1}$, R.~M.~Shannon$^8$, N.~Wex$^1$\\
  $^1$Max-Planck-Institut f\"{u}r Radioastronomie, 
      Auf dem H\"{u}gel 69, D-53121 Bonn, Germany \\
  $^2$Centre for Astrophysics and Supercomputing, 
      Swinburne University of Technology, Mail H39, 
      PO Box 218, VIC 3122, Australia \\
  $^3$The ARC Centre of Excellence for All-Sky 
      Astrophysics (CAASTRO) \\
  $^4$Department of Physics, West Virginia University, 
      Morgantown, WV 26506, USA \\
  $^5$International Centre for Radio Astronomy Research, 
      Curtin University, Bentley, WA 6102, Australia \\    
  $^6$Osservatorio Astronomico di Cagliari, Loc. 
      Poggio dei Pini, Strada 54, 09012 Capoterra
      (CA), Italy \\ 
  $^7$NASA Jet Propulsion Laboratory, M/S 138-307, 
      Pasadena CA 91106, USA \\
  $^8$CSIRO Astronomy $\&$ Space Science, 
      Australia Telescope National Facility, 
      PO Box 76, Epping, NSW 1710, Australia \\
  $^9$Jodrell Bank Centre for Astrophysics, 
      University of Manchester, Alan Turing Building, 
      Oxford Road, Manchester M13 9PL, United Kingdom \\
  $^{10}$Dipartimento di Fisica, Universit\`a di Cagliari, 
      Cittadella Universitaria 09042 Monserrato (CA), Italy \\    
  $^{11}$Laboratoire de Physique et Chimie de l'Environnement 
      et de l'Espace -- Universit\'{e} d'Orl\'{e}ans / CNRS, 
      F-45071 Orl\'{e}ans Cedex 02, France \\
  $^{12}$Argelander-Insitut f\"ur Astronomie, Universit\"at Bonn,Auf dem H\"ugel 71, 53121 Bonn, Germany}

\date{Released 2013 Xxxxx XX}

\pagerange{\pageref{firstpage}--\pageref{lastpage}} \pubyear{2013}

\def\LaTeX{L\kern-.36em\raise.3ex\hbox{a}\kern-.15em
    T\kern-.1667em\lower.7ex\hbox{E}\kern-.125emX}

\begin{document}

\label{firstpage}

\maketitle

\begin{abstract}
We report on the discovery of four millisecond pulsars (MSPs) in the High Time Resolution Universe (HTRU) pulsar survey being conducted at the Parkes 64-m radio telescope. All four MSPs are in binary systems and are likely to have white dwarf companions. In addition, we present updated timing solutions for 12 previously published HTRU MSPs, revealing new observational parameters such as five proper motion measurements and significant temporal dispersion measure variations in PSR~J1017$-$7156. We discuss the case of PSR~J1801$-$3210, which shows no significant period derivative $(\dot{P})$ after four years of timing data. Our best-fit solution shows a $\dot{P}$ of the order of $10^{-23}$, an extremely small number compared to that of a typical MSP. However, it is likely that the pulsar lies beyond the Galactic Centre, and an unremarkable intrinsic $\dot{P}$ is reduced to close to zero by the Galactic potential acceleration. Furthermore, we highlight the potential to employ PSR~J1801$-$3210 in the strong equivalence principle test due to its wide and circular orbit. In a broader comparison with the known MSP population, we suggest a correlation between higher mass functions and the presence of eclipses in `very low-mass binary pulsars', implying that eclipses are observed in systems with high orbital inclinations. We also suggest that the distribution of the total mass of binary systems is inversely-related to the Galactic height distribution. Finally, we report on the first detection of PSRs~J1543$-$5149 and J1811$-$2404 as gamma-ray pulsars. 

\end{abstract}


\section{Introduction}
The bulk of the known pulsar population falls into two distinct groups when plotted on a period-period derivative diagram ($P$-$\dot{P}$~diagram). The normal or slow pulsars typically have spin periods between 0.1 and a few seconds, and have derived surface magnetic field strengths of $10^{11}$ to $10^{13}\rm\,G$. The second group has much lower magnetic field strengths of $10^{8}$ to $10^{9}\rm\,G$ and rapid spin periods measured in milliseconds. Members of this latter population are often referred to as the millisecond pulsars or MSPs. It is believed that MSPs are formed in binary systems in which the pulsar accretes matter from a companion star, gaining mass and angular momentum during the accretion process \citep[e.g.,][]{Alpar1982,Tauris2006}. The pulsar is thus recycled and spun up to a very short spin period. At the same time the strength of its magnetic field is reduced, resulting in the typically small observed period derivative \citep[e.g.,][]{Bhattacharya2002}. This formation scenario holds for most of the Galactic-field MSPs, whereas MSPs found in globular Clusters (GC) have more complicated histories, due to the significant probability of multiple exchange interactions with other cluster stars. In this paper we focus only on MSPs in the Galactic field.

MSPs are of particular interest mainly because of their typically high rotational stability, which combined with their short spin periods and narrow profile features, enable them to be timed precisely. This is in contrast to the younger group of normal or slow pulsars, which often show timing noise manifested as quasi-random variations in the rotational behaviour. MSPs are thus reliable and precise timing tools for a variety of astrophysical applications. For instance, MSPs have been employed in tests of gravity \citep[e.g.,][]{Stairs2003,Freire2012}, for the detection of low frequency gravitational waves in Pulsar Timing Arrays \citep[PTAs;][]{Yardley2011,vHaasteren2011}, to provide measurements of neutron star masses to constrain the Equation of State \citep{Demorest2010,Antoniadis2013}, as precise clocks \citep{Hobbs2012}, in an array to constrain the Solar System ephemeris \citep{Champion2010}, and as an aid for the folding of gamma-ray photons to study the high-energy emission mechanism of pulsars \citep[e.g.,][]{Abdo2009,Espinoza2013}. At the same time, unique systems are constantly being discovered, including triple systems \citep{Lynch2013}, a highly-eccentric system \citep{Champion2008} and the MSP~J1719$-$1438 with an ultra-low mass companion \citep{Bailes2011}, challenging our theories of MSP formation and binary evolution.

To discover more MSPs and to improve our understanding of the MSP population as a whole, we began the High Time Resolution Universe (HTRU) survey in 2008. The HTRU is a blind pulsar survey of the Southern sky with the 64-m Parkes telescope \citep{HTRU1} complemented by a twin survey in the north with the 100-m Effelsberg radio telescope \citep[HTRU-North;][]{Barr2013}. The surveys have benefited from recent advancements in technology and provide unprecedented time and frequency resolution, making them more sensitive to MSPs than previous efforts at these two telescopes. To date, the HTRU survey at Parkes has discovered more than 150 pulsars, of which 27 are MSPs. 

The discovery of pulsars is however just a first step and, in fact, interesting science can usually only be revealed when a follow-up timing campaign is carried out. For MSPs, a coherent timing solution (i.e. when the number of rotations between every observation is well-determined) can be achieved typically within a few weeks of intense timing observations, providing preliminary determination of the rotational and orbital parameters, if any, of the MSP. This is the case for four newly-discovered MSPs presented in this paper. On the other hand, a timing campaign with a longer time baseline is necessary for improving the uncertainties of the timing solution and uncovering subtle details of each MSP system, such as proper motion, parallax, and possibly post-Keplerian binary parameters. This is demonstrated here by the further timing of 12 HTRU MSPs, the discoveries of which were first published two years ago \citep{HTRU2,HTRU4,Bailes2011}. 

In this paper we describe the observations and analysis procedures used for obtaining the timing solutions (Section~\ref{sec:obs}). We report on the discovery of four MSPs and present their initial timing solutions in Section~\ref{sec:4new}, which include a discussion on the nature of the binary companions. We present the updated timing parameters for 12 further MSPs in Section~\ref{sec:12MSPs}, followed by a detailed discussion on various scientific implications arising from our measurements. Finally in Section~\ref{sec:conclusion} we present our conclusions. 

\section{OBSERVATIONS AND ANALYSIS} \label{sec:obs}
\begin{table}
\setlength{\tabcolsep}{0.13cm}
 \caption{Specifications of the observing system employed for the timing observations in this work. $G$ represents the antenna gain and $T_{\rm{sys}}$ is the receiver system temperature. $f_{\rm{c}}$ represents the central frequency in MHz and $B$ is the bandwidth in MHz.}
 \begin{minipage}{9cm}
\begin{tabular}{lp{1.1cm}p{0.6cm}lp{0.8cm}p{0.9cm}}
  \hline
  Receiver  & $G$ $\rm(K\rm\,Jy^{-1})$ & $T_{\rm{sys}}$ (K) & Backend       & $f_{\rm{c}}$ (MHz) & $B$ (MHz)  \\
  \hline 
  10/50CM   & 0.74  & 40           & Parkes DFBs   & 732                & 64         \\
            &   &               & Parkes APSR   & 732                & 64        \\
            &   &               & Parkes CASPSR & 728                & 64        \\
  \hline
  Multibeam & 0.74 & 23 & Parkes DFBs   & 1369               & 256        \\
            &       &    & Parkes BPSR   & 1382               & 300        \\
            &       &    & Parkes APSR   & 1369               & 256       \\
            &       &    & Parkes CASPSR & 1382               & 320$^{\dagger}$  \\  
  Single-pixel & 1.00   & 28   & Jodrell DFB  & 1532                & 384        \\
            &    &    & Jodrell ROACH & 1532               & 400       \\
  \hline
  10/50CM   & 0.74   & 30  & Parkes DFBs   &  3094           & 900        \\ 
  \hline \label{tab:specs}
 \end{tabular}
\vspace{-0.2\skip\footins}
 \begin{flushleft}
 $^{\dagger}$ CASPSR has a bandwidth of 400\,MHz, but only 320\,MHz can be used due to the Thuraya-3 filters. \\
 \end{flushleft}

 \end{minipage}
\end{table}

All 16 MSPs presented in this paper were discovered in 540-s-long integrations as part of the medium-latitude section ($-120\degr<\textit{l}<30\degr$, $|b|<15\degr$) of the HTRU survey. Survey observations were made using the 13-beam Multibeam receiver \citep{Multibeam1996} on the 64-m Parkes radio telescope. Full details of the survey parameters are given in \citet{HTRU1}. 

Follow-up timing observations were made at Parkes initially with a setup similar to that of the survey, employing the central beam of the same 13-beam Multibeam receiver at a centre frequency near 1.4\,GHz and the Berkeley-Parkes-Swinburne Recorder (BPSR) with 1024 frequency channels incoherently dedispersed at a time resolution of 64$\mu$s. Later when the pulsar parameters were identified with sufficient accuracy, observations were carried out using the Digital Filterbank systems (DFBs) which are based on the implementation of a polyphase filter in FPGA processors with incoherent dedispersion. Coherently dedispersed data are collected by the CASPER Parkes Swinburne Recorder\footnote{http://astronomy.swin.edu.au/pulsar/?topic=caspsr} (CASPSR) and the ATNF Parkes Swinburne Recorder\footnote{http://astronomy.swin.edu.au/pulsar/?topic=apsr} (APSR). Pulsars with declination above $-35\degr$ are also being timed at the Jodrell Bank Observatory with the Lovell 76-m telescope, using a DFB backend and a ROACH backend. The latter is based on the ROACH FPGA processing board\footnote{https://casper.berkeley.edu/wiki/ROACH} and coherently dedisperses the data. Refer to Table~\ref{tab:specs} for the specifications of all observing systems employed. 

Observations have also been taken at different frequencies at Parkes using the 10/50\,cm receiver \citep{1050CM}, to allow for precise dispersion measure (DM) measurements and to study any variations of pulsar profiles across frequencies. The various combinations of receivers and backends had central frequencies as listed in column 3 of Table \ref{tab:specs}. Note that predetermined offsets were applied to the observational data from Parkes to account for instrumental delay across observations with different backends in accordance with \citet{Manchester2013}.

Timing observations of these 16 pulsars have first been made with an intense timing campaign within roughly their first year of discovery, and gradually decreased to weekly observation for the case of Jodrell Bank observations, whereas Parkes observations are more irregular with gaps ranging from days to months depending on telescope availability. Integration times vary from a few minutes to more than 2 hours, with longer observations for weaker pulsars to achieve adequate signal-to-noise (S/N) of at least 10.

We have used the \textsc{psrchive} data analysis package \citep{Hotan2004} for data reduction. Each observation is corrected for dispersion and folded at the predicted topocentric pulse period, before finally summing over both frequency and time to produce an integrated profile. We align these profiles from each observation using an ephemeris created from the initial timing solution. This forms the basis of a noise-free analytic reference template, and we convolve the template with each individual profile to produce a Time of Arrival (TOA) \citep{Taylor1992}. The DE421 Solar System ephemeris of the Jet Propulsion Laboratory \citep{DE421} was used to transform the TOAs to the Solar System barycentre. The \textsc{tempo2} software package \citep{Hobbs2006} was then used to fit a timing model to all TOAs, taking into account the astrometry, spin, and orbital motion of the pulsar. This process of cross-correlating a template with individual profiles can then be iterated to improve the quality of the model fit. We generate multiple TOAs per observation when possible, especially for the pulsars with small orbital periods. This is to make sure each TOA does not cover more than one tenth of an orbit, to avoid masking orbital information within a seemingly high S/N TOA. If simultaneous observations with different backends were taken, we include only one of the observations to avoid otherwise over-weighting duplicated TOAs.  

All 16 MSPs in this work are in binary systems. The Damour-Deruelle (DD) timing model \citep{DD1986} in \textsc{tempo2} is a theory-independent description for eccentric binary orbits. However, for binaries with small eccentricities the location of periastron is not well-defined and using the DD timing model results in a high covariance between the longitude of periastron $(\omega)$ and the epoch of periastron $\rm(T_{0})$. A useful quantity to help choosing the best timing model is $xe^2$, where $e$ is the eccentricity and $x$ is the projected semi-major axis of the pulsar orbit as defined by:
\begin{equation}
x \equiv \frac{a_{\rm{p}}\sin{i}}{c}\,,
\label{eq:x}
\end{equation}
with $a_{\rm{p}}$ being the semi-major axis, $i$ the orbital inclination and $c$ the speed of light. For pulsars with $xe^2$ smaller than the timing precision as represented by the RMS, we use the ELL1 timing model \citep{Lange2001} alternatively. The ELL1 timing model avoids the covariance by using the Laplace-Lagrange parameters ($\epsilon_{1}=e\sin\omega$ and $\epsilon_{2}=e\cos\omega$) and the time of ascending node passage $(\rm{T_{asc}})$ instead of $\rm{T_{0}}$ as in the DD timing model. 

Towards the end of the timing analysis procedure when the respective reduced $\rm\chi^2$ comes close to one, we can then assume a reliable fit is achieved which is only influenced by the presence of radiometer noise in the template. As a last step, we compensate for these systematic effects by calculating dataset-specific calibration coefficients (also known as `EFAC' in {\textsc{tempo2}). These coefficients are applied to scale the TOA uncertainties such that each final respective reduced $\rm\chi^2$ is unity. 

In addition, full flux density and polarisation calibration are implemented for the four newly-discovered MSPs, in order to study their polarisation profiles. This analysis is not repeated for the rest of the 12 MSPs in this paper since their polarisation properties are already presented in \citet{HTRU4}. With the only exception of PSR~J1017$-$7156, a high-precision timing pulsar which is noticeably polarised in both linear and circular sense, we have fully calibrated the data to correctly assess the uncertainties on the TOAs. 

To carry out the calibration we make use of Parkes DFB observations which record the four Stokes parameters in each frequency channel. We calibrate each observation for the differential gain and phase between the feed with an observation of the noise diode coupled to the receptors in the feeds. This calibration observation triggers a square-wave signal which is used to retrieve the true Stokes parameters, and it is important that this calibration is taken adjacent to the targeted pulsar observations. In addition, we correct for the non-orthogonality of the receptors in the Multibeam receiver by computing a model of the Jones matrix for the receiver using an averaged observation of the bright pulsar J0437$-$4715, in accordance with the `measurement equation modelling' technique described in \citet{vStraten2004}, and we calibrate the flux density by using an averaged observation of Hydra A. 

\section{DISCOVERY OF FOUR MILLISECOND PULSARS} \label{sec:4new}
We present the discoveries of four MSPs in the HTRU survey, namely PSRs~J1056$-$7117, J1525$-$5545, J1528$-$3828 and J1755$-$3716. They all have observations spanning more than one year, and their coherent timing solutions are shown in Table~\ref{tab:solutionnew}. All four are in binary systems. 

\subsection{On the nature of the binary companions} \label{4newBincomp}
PSR~J1528$-$3828 and PSR~J1056$-$7117 are likely to be formed from wide-orbit low mass X-ray binaries (LMXBs), leading to the formation of classic MSPs with Helium white dwarf (He-WD) companions. According to \citet{Tauris2011}, wide-orbit LMXBs with $P_{\rm{orb}}\ge1$\,day lead to He-WDs with masses between about $0.15$ to $0.46\,M_{\sun}$.

PSR~J1755$-$3716 has a relatively high median companion mass of $0.35\,M_{\sun}$. Although this would fit in the above classification, the fact that PSR~J1755$-$3716 has a spin period of 12.8\,ms implies that the system is only mildly recycled. This, combined with its $P_{\rm{orb}}$ of just 11.5\,days (which is too short for LMXB evolution to produce a $0.35\,M_{\sun}$ WD, \citet{Tauris1999}), indicates that its evolutionary track is more likely to have started from an intermediate mass X-ray binary pulsar (IMXB) accreting via early Case B Roche-lobe overflow (RLO) \citep{Tauris2011}. The companion of PSR~J1755$-$3716 is probably a CO-WD.

PSR~J1525$-$5545 has a solar mass companion with a median mass of 0.99\,$M_{\sun}$ and an $P_{\rm{orb}}$ of 0.99\,days. These fit the typical characteristics of binary evolution from a wide-orbit IMXB via Case C RLO and common envelope evolution \citep{Tauris2011}. The companion is likely to be a massive CO-WD, or an ONeMg-WD if the orbital inclination is low.

\subsection{Polarisation Profiles} \label{sec:polarisation}
\begin{figure*}
\centering
\setlength\fboxsep{0pt}
\setlength\fboxrule{0pt}
\fbox{\includegraphics[width=15cm]{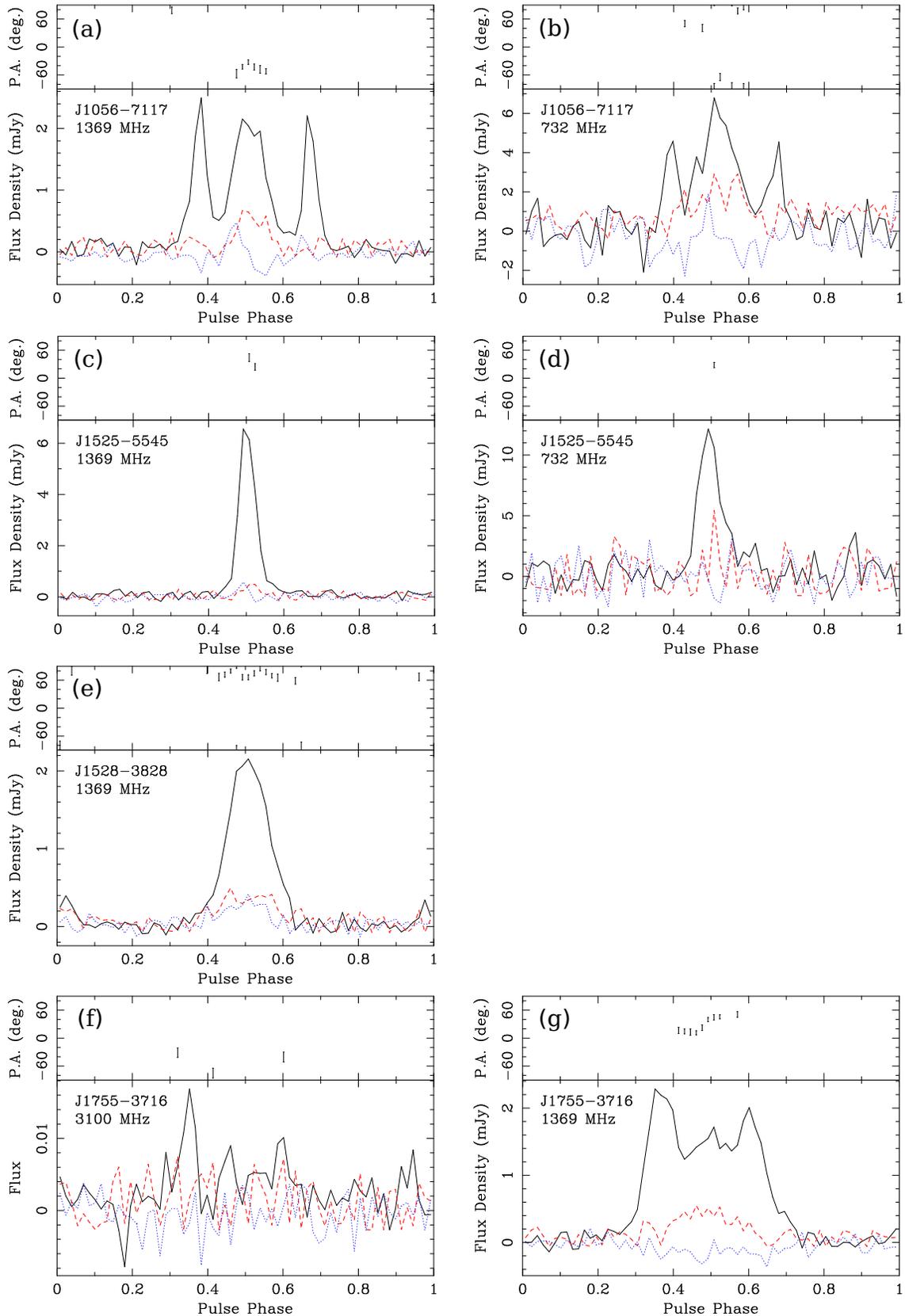}}
  \caption{Polarisation profiles of PSRs~J1056$-$7117 at (a) 1369 and (b) 732\,MHz, J1525$-$5545 at (c) 1369 and (d) 732\,MHz, J1528$-$3828 at (d) 1369\,MHz, and J1755$-$3716 at (e) 3100 and (f) 1369\,MHz. The upper panel shows the RM-corrected P.A. variation in longitude with respect to the celestial north. The lower panel shows the integrated profile where the black solid line, red dashed line and blue dotted line represent total intensity, linear and circular polarisation respectively.}
  \label{fig:polarisation}
\end{figure*}

Figure~\ref{fig:polarisation} shows the integrated polarisation profiles of the four MSPs in total intensity, linear and circular polarisation. We measure the Faraday rotation observed towards each pulsar by fitting the position angle (P.A.) variations across the 256\,MHz band centred at 1369\,MHz, and the plots shown here have their rotation measure (RM) corrected with the respective RMs as listed in Table~\ref{tab:solutionnew}. Multi-frequency data are included only if the S/N ratio is high enough, and are plotted here with an arbitrary alignment. None of the four MSPs are detectable at 3100\,MHz with at least 1 hour of observation, except a tentative detection of PSR~J1755$-$3716. At 732\,MHz only PSRs~J1056$-$7117 and J1525$-$5545 are detectable, both with limited S/N. Although pulsars typically have steep spectral indexes and thus higher flux at lower observing frequencies, our receiver system at 50\,cm has a reduced sensitivity due to its higher system temperature and narrow bandwidth (Table~\ref{tab:specs}). Hence we cannot comment if there is any profile evolution across frequency.  

PSR~J1056$-$7117 has a profile comprising three components. The emission of the middle component changes handedness in circular polarisation, whereas the S/N of the other two components are not sufficient for identifying the polarisation fraction. Linear polarisation is present in the middle component, although noisy. PSR~J1525$-$5545 has a simple, single peak profile. It is almost completely unpolarised, and such low polarisation profile is typically associated with aligned gamma-ray and radio profiles \citep{Espinoza2013}. Although no \textit{Fermi Gamma-ray Space Telescope (Fermi)} association has been reported for PSR~J1525$-$5545 yet, it is worth following-up as the radio ephemeris improves with longer timing baseline. PSR~J1528$-$3828 has a broad single peak profile with a hint of interpulse, and the P.A. is relatively flat over the profile. PSR~J1755$-$3716 also has a profile formed of three components with some degree of linear polarisation in the middle component which is narrower compared to the total intensity, and the P.A. seems to show an `S-shaped' swing.  

\begin{table*}
  \begin{minipage}{18cm}
    \centering
    \caption{\textsc{tempo2} best-fit parameters for the four newly-discovered MSPs. Values in parentheses are the nominal 1-$\sigma$ uncertainties in the last digits. The last panel shows derived parameters, the respective equations for which can be found in \protect\citet{Handbook2004}, except for the DM distance which is derived according to \protect\citet{NE2001model}.} 
    \begin{tabular}{P{5.8cm}llll}
  \hline
  Parameter & J1056$-$7117 & J1525$-$5545 & J1528$-$3828 & J1755$-$3716 \\
  \hline
  Right ascension, $\alpha$ (J2000) & 10:56:45.980(4) & 15:25:28.1340(2)   & 15:29:15.1066(10)& 17:55:35.4462(4)\\
  Declination, $\delta$ (J2000)     & $-$71:17:53.394(14) & $-$55:45:49.842(5) & $-$38:28:45.85(3)   & $-$37:16:10.78(4) \\
  Galactic longitude, $l$ $(\degr)$ & 293.933 & 323.439 & 333.886 & 353.882\\
  Galactic latitude, $b$ $(\degr)$ & $-$10.458 & 0.851 & 14.728 & $-$6.041 \\
  Spin frequency, $\nu$ (Hz) & 38.0088284880(10)& 88.02908501431(14)& 117.8372326493(7) & 78.2101189443(6)\\
  Spin period, $P$ (ms) & 26.3096769823(7) & 11.359881791766(18) & 8.48628211573(5) & 12.78606928998(9)\\
  Frequency derivative, $\dot{\nu}$ $\rm(s^{-2})$ & $-$9.1(9)$\times10^{-17}$ & $-$1.018(4)$\times10^{-15}$ & $-$3.75(18)$\times10^{-16}$&$-$1.9(2)$\times10^{-16}$ \\
  Period derivative, $\dot{P}$ & 6.3(6)$\times10^{-20}$ & 1.313(5)$\times10^{-19}$ & 2.70(13)$\times10^{-20}$ & 3.1(3)$\times10^{-20}$\\
  Dispersion measure, DM $\rm(cm^{-3}\rm\,pc)$ &  93.04(4) & 126.934(7) & 73.62(2) & 167.603(19) \\
  \hline
  Orbital period, $P_{\rm{orb}}$ (days)& 9.1387994(5) & 0.9903149542(7) & 119.674809(16) & 11.5156057(3)\\ 
  Projected semi-major axis, $x$ (lt-s)& 4.14855(2)  & 4.710520(6) & 29.34054(2)  & 10.645131(12)\\
  Epoch of ascending node, $T_{\rm{asc}}$ (MJD)& 57436.53532(7) & 55891.5285616(2) & 55941.60(16)$^{\dagger}$    & 55958.790341(9)\\
  $e\sin\omega$, $\epsilon_{1}$ $(10^{-6})$ & 6(8) & $-$4.4(17) & $-\,^{\dagger}$ & $-$7(2)\\
  $e\cos\omega$, $\epsilon_{2}$ $(10^{-6})$ & $-$12(10)& $-$1.8(16) & $-\,^{\dagger}$ & 12(3)\\
  Inferred eccentricity, $e$ $(10^{-6})$ & 14(10)& 4.8(17) & 168.6(14)$^{\dagger}$            & 14(3)\\
  Longitude of periastron, $\omega$ $(\degr)$ & 150(30) & 247(19) & 282.2(4) & 329(9)\\
  Minimum companion mass$^{*}$, $m_{\rm{c,min}}$ $(M_{\sun})$& 0.13 &  0.81 & 0.16& 0.30 \\
  Median companion mass$^{**}$, $m_{\rm{c,med}}$ $(M_{\sun})$& 0.15 & 0.99 & 0.19 & 0.35 \\
  \hline
  Binary model & ELL1 & ELL1 & DD & ELL1 \\
  First TOA (MJD) & 55954.5 & 55987.6 & 55905.0 & 56053.9 \\
  Last TOA (MJD)   & 56491.1 & 56510.5 & 56510.5 & 56510.6 \\
  Timing epoch (MJD) & 57436.5 & 55891.5 & 55847.0 & 55958.8 \\
  Points in fit      & 24     & 25       & 31       & 27 \\
  Weighted RMS residuals ($\mu$s) & 41 & 8.3 & 51 & 19 \\
  Reduced $\chi^{2}\,^{\ddagger}$   & 0.9 & 0.9 & 2.0 & 0.7 \\
\hline
  Mean flux density at 1.4-GHz, $S_{1400}$ (mJy) & 0.34 & 0.33 & 0.16 & 0.53 \\
  Pulse width at 50$\%$ of peak, $W_{50}$ ($\degr$) & 69 & 17 & 52 & 110 \\
  Rotation measure, RM $\rm(rad\rm\,m^{-2})$ & $-$22(8) & $-$19(9) & $-$29(9) & 54(3) \\
\hline
  DM distance (kpc)  & 2.6 & 2.4 & 2.2 & 3.9 \\
  Characteristic age, $\tau_{\rm{c}}$ (Myr) & 6.6$\times10^{3}$ & 1.4$\times10^{3}$ & 5.0$\times10^{3}$ & 6.4$\times10^{3}$ \\
  Spin down energy loss rate, $\dot{E}$ $(10^{33}\rm\,erg\rm\, s^{-1})$  & 0.14 & 3.5 & 1.7 & 0.57 \\
  $\dot{E}/d^{2}$ $(10^{33}\rm\,erg\rm\,kpc^{-2}\rm\,s^{-1})$  & 0.021 & 0.62 & 0.36 & 0.037 \\
  Characteristic dipole surface magnetic field strength at equator, $B_{\rm{eq}}$ ($10^{8}$\,G) & 13 & 12 & 4.9 & 6.5 \\
 \hline \label{tab:solutionnew}
 \end{tabular}
\vspace{-0.5\skip\footins}
 \begin{flushleft}
 $^{*}$ $m_{\rm{c,min}}$ is calculated for an orbital inclination of $i=90^{\circ}$ and an assumed pulsar mass of $1.35\,M_{\sun}$. \\
 $^{**}$ $m_{\rm{c,med}}$ is calculated for an orbital inclination of $i=60^{\circ}$ and an assumed pulsar mass of $1.35\,M_{\sun}$. \\
 $^{\dagger}$ For PSR~J1528$-$3228 the DD model is used. We quote $\rm{T_{0}}$ instead of $\rm{T_{asc}}$. $e$ is directly fitted for and not inferred from the $\rm\epsilon$ parameters.  \\
 $^{\ddagger}$ The reduced $\chi^{2}$ stated here represents the value before the application of EFAC. Note that the rest of the timing solutions have EFACs incorporated, bringing the reduced $\chi^{2}$ to unity. \\
 \end{flushleft}
 \end{minipage}
\end{table*}

\begin{landscape}
\begin{table}
\centering
 \caption{\textsc{tempo2} best-fit parameters using the ELL1 timing model. Values in parentheses are the nominal 1-$\sigma$ uncertainties in the last digits. If only an upper limit is constrained, we quote it at the 2-$\sigma$ level. The last panel shows derived parameters, the respective equations for which can be found in \protect\citet{Handbook2004}, except for the DM distance which is derived according to \protect\citet{NE2001model}.}
   \begin{tabular}{lllll}
  \hline
  Parameter & J1337$-$6423 & J1446$-$4701 & J1502$-$6752 & J1543$-$5149 \\
  \hline
  Right ascension, $\alpha$ (J2000) & 13:37:31.883(2)   & 14:46:35.71391(2) & 15:02:18.615(2) & 15:43:44.1498(2)\\
  Declination, $\delta$ (J2000)     & $-$64:23:04.915(9)& $-$47:01:26.7675(4) & $-$67:52:16.759(18) &$-$51:49:54.685(2) \\
  Galactic longitude, $l$ ($\degr$) & 307.889 & 322.500 & 314.798 & 327.920 \\
  Galactic latitude, $b$ ($\degr$) & $-$1.958 & $+$11.425 & $-$8.067 & $+$2.479\\
  Spin frequency, $\nu$ (Hz) & 106.11873496995(19)   & 455.644016442381(13) & 37.39097199147(8) & 486.15423208300(13)\\
  Spin period, $P$ (ms) & 9.423406717796(17)   & 2.19469577985000(6) & 26.74442376699(6)& 2.0569603924156(5)\\
  Frequency derivative, $\dot{\nu}$ $(s^{-2})$ & $-$2.78(2)$\times10^{-16}$ & $-$2.0367(4)$\times10^{-15}$ & $-$4.397(19)$\times10^{-16}$ & $-$3.819(3)$\times10^{-15}$\\
  Period derivative, $\dot{P}$ & 2.47(2)$\times10^{-20}$ & 9.810(2))$\times10^{-21}$  & 3.145(13)$\times10^{-19}$ &1.6161(14)$\times10^{-20}$ \\
  Dispersion measure, DM $\rm(cm^{-3}\rm\,pc)$ & 259.2(13) & 55.83202(14) & 151.2(18) &50.93(14) \\
  Proper motion in $\alpha$, $\mu_{\alpha}$ (mas$\rm\,yr^{-1})$ & $-$6(6)& $-$4.0(2)   & $-$6(9) &$-$4.3(14) \\
  Proper motion in $\delta$, $\mu_{\delta}$ (mas$\rm\,yr^{-1})$ & $-$7(5) & $-$2.0(3)  & $-$14(16) & $-$4(2)\\
  \hline
  Orbital period, $P_{\rm{orb}}$ (days)& 4.785333912(5) & 0.27766607732(13)& 2.48445723(18) &8.060773125(9)\\
  Projected semi-major axis, $x$ (lt-s)& 13.086505(5) & 0.0640118(3) &  0.31754(2)   &6.480288(2) \\
  Epoch of ascending node, $T_{\rm{asc}}$ (MJD)& 55234.7703674(6) & 55647.8044392(2) & 55421.21199(3)& 54929.0678261(11)\\
  $e\sin\omega$, $\epsilon_{1}$ $(10^{-6})$ & 18.3(8)& 18(8) & 21(140) &20.8(5) \\
  $e\cos\omega$, $\epsilon_{2}$ $(10^{-6})$ & 7.7(9)& $-$11(9) & $-$23(150) &5.3(6)\\
  Inferred eccentricity, $e$ $(10^{-6})$ & 19.8(8)& 21(8) & $<$330  & 21.5(5)\\
  Longitude of periastron, $\omega$ $(\degr)$ & 67(2) & 120(20) & 130(260) & 75.6(16)\\
Minimum companion mass$^*$, $m_{\rm{c,min}}$ $(M_{\sun})$& 0.78 & 0.019 & 0.022 &0.22\\
 Median companion mass$^{**}$, $m_{\rm{c,med}}$ $(M_{\sun})$& 0.95 & 0.022 & 0.025 &0.26 \\
\hline
  Binary model & ELL1 & ELL1 & ELL1 & ELL1 \\
  First TOA (MJD) & 55540.0 & 55460.0 & 55360.4 &55540.8 \\
  Last TOA (MJD)   & 56510.2 & 56497.2 & 56510.3 &56510.3 \\
  Timing epoch (MJD) &  55234.7    & 55647.8       & 55421.2  &55522  \\
  Points in fit      & 76     & 154       & 57       &52 \\     
  Weighted RMS residuals ($\mu$s) & 26 & 2.1 & 87 &6.9 \\
  Reduced $\chi^{2}\,^{\ddagger}$  & 0.9 & 1.0 & 1.0 &1.2 \\
\hline
  Mean flux density at 1.4-GHz, $S_{1400}$ (mJy) & 0.29    & 0.40       & 0.69     &0.55  \\
  Pulse width at 50$\%$ of peak, $W_{50}$ ($\degr$) & 28     & 18       & 40       &49     \\
\hline
  DM Distance, $d$ (kpc) & 5.1 & 1.5 & 4.2 & 2.4 \\
  Transverse velocity, $V_{\rm{T}}$ $(\rm km\rm\,s^{-1})$ & 230(140) & 32(8) & $<960$ & 70(30)\\
  Intrinsic period derivative, $\dot{P}_{\rm{int}}$ ($\rm10^{-20}$) & 1.0(13) & 0.972(2)   & 15(14) & 1.54(3) \\
  Characteristic age$^{\dagger\dagger}$, $\tau_{\rm{c}}$ (Myr) & 1.4$\times10^{4}$ & 3.6$\times10^{3}$ & 2.7$\times10^{3}$ &2.1$\times10^{3}$ \\
  Spin down energy loss rate$^{\dagger\dagger}$, $\dot{E}$ $(10^{33}\rm\,erg\rm\, s^{-1})$ & 0.51 & 36 & 0.32 & 70 \\
  $\dot{E}/d^{2}\,^{\dagger\dagger}$ $(10^{33}\rm\,erg\rm\,kpc^{-2}\rm\,s^{-1})$ & 0.020 & 16 & 0.018 & 12 \\
  Characteristic dipole surface magnetic field strength at equator$^{\dagger\dagger}$, $B_{\rm{eq}}$ ($10^{8}$\,G) & 3.2 & 1.5 & 21 &1.8 \\
 \hline \label{tab:ell1}
 \end{tabular}%} 

\vspace{-0.3\skip\footins}                                                                                                                                                                       
 \begin{flushleft}
 $^*$ $m_{\rm{c,min}}$ is calculated for an orbital inclination of $i=90^{\circ}$ and an assumed pulsar mass of $1.35\,M_{\sun}$. \\
 $^{**}$ $m_{\rm{c,med}}$ is calculated for an orbital inclination of $i=60^{\circ}$ and an assumed pulsar mass of $1.35\,M_{\sun}$. \\
 $^{\ddagger}$ The reduced $\chi^{2}$ stated here represents the value before the application of EFAC. Note that the rest of the timing solutions have EFACs incorporated, bringing the reduced $\chi^{2}$ to unity. \\
 $^{\dagger\dagger}$ These parameters are derived from the intrinsic period derivatives $\dot{P}_{\rm{int}}$. For the derivation of $\dot{P}_{\rm{int}}$ refer to Section~\ref{sec:pdot}.\\
 \end{flushleft}
\end{table}
\end{landscape}

\begin{landscape}
\begin{table}
\centering
 \caption{\textsc{tempo2} best-fit parameters using the ELL1 timing model. Values in parentheses are the nominal 1-$\sigma$ uncertainties in the last digits. The last panel shows derived parameters, the respective equations for which can be found in \protect\citet{Handbook2004}, except for the DM distance which is derived according to \protect\citet{NE2001model}.}
   \begin{tabular}{lllll}
  \hline
  Parameter  & J1622$-$6617 & J1719$-$1438 & J1801$-$3210 & J1811$-$2405\\
  \hline     
  Right ascension, $\alpha$ (J2000) & 16:22:03.6681(4) & 17:19:10.07293(5) & 18:01:25.8896(2) & 18:11:19.85315(2)\\
   Declination, $\delta$ (J2000)     & $-$66:17:16.978(6) & $-$14:38:00.942(4) & $-$32:10:53.714(17) & $-$24:05:18.365(11)\\
  Galactic longitude, $l$ ($\degr$) &  321.977 & 8.858 & 358.922 & 7.073   \\   
  Galactic latitude, $b$ ($\degr$)  & $-$11.56 & $+$12.838 & $-$4.577 & $-$2.559   \\   
   Spin frequency, $\nu$ (Hz)        & 42.33082901464(2)  & 172.707044602370(13)& 134.16363857901(4) & 375.856014397575(9)\\ 
   Spin period, $P$ (ms)             & 23.623444739389(12) & 5.7901517700238(4)& 7.453584373467(2) & 2.66059331683918(7)\\
   Frequency derivative, $\dot{\nu}$ $(s^{-2})$        & $-$1.054(4)$\times10^{-16}$ & $-$2.399(2)$\times10^{-16}$ & 8(7)$\times10^{-19}$ & $-$1.8898(2)$\times10^{-15}$\\ 
   Period derivative, $\dot{P}$ &  5.88(2)$\times10^{-20}$   & 8.044(8)$\times10^{-21}$  & $-$4(4)$\times10^{-23}$  & 1.33780(16)$\times10^{-20}$ \\
   Dispersion measure, DM $\rm(cm^{-3}\rm\,pc)$        & 88.024(9) & 36.862(4) & 177.713(4) & 60.6005(17)\\   
   Proper motion in $\alpha$, $\mu_{\alpha}$ (mas$\rm\,yr^{-1})$ & $-$3(2)   & 1.9(4) & $-$8(2)    & 0.53(13)\\
   Proper motion in $\delta$, $\mu_{\delta}$ (mas$\rm\,yr^{-1})$ & $-$6(4)   & $-$11(2)  & $-$11(10)& $-\,^{\P}$ \\
   \hline
   Orbital period, $P_{\rm{orb}}$ (days)& 1.640635150(8)& 0.0907062900(12)& 20.77169942(8)  & 6.2723020692(12)\\
   Projected semi-major axis, $x$ (lt-s)&0.979386(5) &0.0018212(7) &  7.809317(4) & 5.7056616(3)\\  
   Epoch of ascending node, $T_{\rm{asc}}$ (MJD)& 55253.087283(2)   & 55235.516505(8)& 55001.934484(2) & 55136.16862345(7)\\
   $e\sin\omega$, $\epsilon_{1}$ $(10^{-6})$  & $-$4(12) & $-$700(700) & 1.7(11) & 1.46(10)\\ 
   $e\cos\omega$, $\epsilon_{2}$ $(10^{-6})$ &  14(11)   & 400(700) & 1.0(10)& 0.75(10)\\
   Inferred eccentricity, $e$ $(10^{-6})$      & 14(11)& 800(700)  & 2.0(11) & 1.64(10) \\
   Longitude of periastron, $\omega$ $(\degr)$ & 340(40) & 300(50)& 50(30) & 62(3)\\
   Minimum companion mass$^*$, $m_{\rm{c,min}}$ $(M_{\sun})$& 0.092 & 0.0011 & 0.14 & 0.23\\   
   Median companion mass$^{**}$, $m_{\rm{c,med}}$ $(M_{\sun})$& 0.11 & 0.0013 & 0.16 & 0.27 \\   
   \hline
   Binary model & ELL1 & ELL1 & ELL1 & ELL1 \\
   First TOA (MJD)      & 55256.9 & 55237.0 & 54996.4 & 55136.1\\   
   Last TOA (MJD)     & 56510.3 & 56491.6 & 56485.7 & 56411.2\\   
  Timing epoch (MJD) & 55253.1 & 55235.5 & 55001.9 & 55208.5 \\
  Points in fit      & 86     & 236 & 135     & 97 \\
  Weighted RMS residuals ($\mu$s) & 31 & 10 & 37 & 2.8 \\   
  Reduced $\chi^{2}\,^{\ddagger}$ & 1.3 & 2.1 & 1.6 & 2.6 \\   
\hline
  Mean flux density at 1.4-GHz, $S_{1400}$ (mJy)    & 0.60   & 0.42& 0.32 & 0.37 \\
  Pulse width at 50$\%$ of peak, $W_{50}$ ($\degr$) &  12    & 19  & 30 & 16 \\
\hline
  DM Distance, $d$ (kpc)  & 2.2 & 1.2 &  4.0 & 1.8   \\   
  Transverse velocity, $V_{\rm{T}}$ $\rm(km\,s^{-1})$ & 40(20)  & 60(20) & 270(170) & $-\,^{\P}$\\
  Intrinsic period derivative, $\dot{P}_{\rm{int}}$ $(10^{-20})$ & 5.0(8)   & 0.54(11) & $-$2.7(17)$\,^{\dagger}$ & 1.284(15)$\,^{\P}$ \\ 
  Characteristic age$^{\dagger\dagger}$, $\tau_{\rm{c}}$ (Myr) & 7.5$\times10^{3}$& 1.7$\times10^{4}$ & $>$1.5$\times10^{4}\,^{\dagger}$ & 3.3$\times10^{3}$  \\ 
  Spin down energy loss rate$^{\dagger\dagger}$, $\dot{E}$ $(10^{33}\rm\,erg\rm\, s^{-1})$ & 0.15 & 1.1 & $<0.78\,^{\dagger}$ & 27 \\
  $\dot{E}/d^{2}\,^{\dagger\dagger}$ $(10^{33}\rm\,erg\rm\,kpc^{-2}\rm\,s^{-1})$ & 0.031 & 0.76 & $<0.048\,^{\dagger}$ & 8.3 \\
  Characteristic dipole surface magnetic field strength at equator$^{\dagger\dagger}$, $B_{\rm{eq}}$ ($10^{8}$\,G) & 11 & 1.8 & $<2.5\,^{\dagger}$ & 1.9  \\   
 \hline \label{tab:ell2}
 \end{tabular}
 \begin{flushleft}
 $^*$ $m_{\rm{c,min}}$ is calculated for an orbital inclination of $i=90^{\circ}$ and an assumed pulsar mass of $1.35\,M_{\sun}$. \\
 $^{**}$ $m_{\rm{c,med}}$ is calculated for an orbital inclination of $i=60^{\circ}$ and an assumed pulsar mass of $1.35\,M_{\sun}$. \\
 $^{\ddagger}$ The reduced $\chi^{2}$ stated here represents the value before the application of EFAC. Note that the rest of the timing solutions have EFACs incorporated, bringing the reduced $\chi^{2}$ to unity. \\
 $^{\dagger\dagger}$ These parameters are derived from the intrinsic period derivatives $\dot{P}_{\rm{int}}$. For the derivation of  $\dot{P}_{\rm{int}}$ refer to Section~\ref{sec:pdot}.\\
 $^{\dagger}$ For PSR~J1801$-$3210 the potential causes of this apparent negative $\dot{P}_{\rm{int}}$ is discussed in Section~\ref{sec:pdot1801}. The period derivative related parameters are derived with the 2-$\sigma$ upper limit of $\dot{P}_{\rm{int}} < 8.1\times10^{-21}$. \\ 
   $^{\P}$ For PSR~J1811$-$2405 we have fixed the unconstrained $\mu_{\delta}$ at zero because this pulsar is very close to the ecliptic plane. Its $V_{\rm{T}}$ is therefore also not measurable. The derived $\dot{P}_{\rm{int}}$ only symbolises a lower limit without correcting for any Shklovskii contribution in $\mu_{\delta}$.\\
 \end{flushleft}
\end{table} 
\end{landscape}

\begin{landscape}
\begin{table}
\centering
\caption{\textsc{tempo2} best-fit parameters using the DD timing model, except in the case of PSR~J1731$-$1847, for which we have instead used BTX model to accommodate the higher order orbital period changes. Values in parentheses are the nominal 1-$\sigma$ uncertainties in the last digits. The fifth panel shows derived parameters, the respective equations for which can be found in \protect\citet{Handbook2004}, except for the DM distance which is derived according to \protect\citet{NE2001model}.}
\begin{tabular}{lllll}
  \hline
  Parameter & J1017$-$7156 & J1125$-$5825 &  J1708$-$3506 & J1731$-$1847 \\
  \hline
   Right ascension, $\alpha$ (J2000) & 10:17:51.32828(2)& 11:25:44.36564(5) & 17:08:17.62215(10) & 17:31:17.609823(17) \\
   Declination, $\delta$ (J2000) & $-$71:56:41.64586(11)& $-$58:25:16.8798(4)& $-$35:06:22.640(4) & $-$18:47:32.666(3)\\
  Galactic longitude, $l$ ($\degr$) & 291.558 & 291.893 & 350.469 &  6.880 \\
  Galactic latitude, $b$ ($\degr$) & $-$12.55 & $+$2.602 & $+$3.124 &  $+$8.151 \\
  Spin frequency, $\nu$ (Hz)  & 427.621905105409(6)  & 322.350432991279(16) & 221.96775106948(3)  & 426.51934403983(2)\\
  Spin period, $P$ (ms) &  2.33851444011854(3)  & 3.10221391893416(16) & 4.5051589484588(6)  & 2.34455954688563(11) \\
  Frequency derivative, $\dot{\nu}\,\rm(s^{-2})$  & $-$4.0584(12)$\times10^{-16}$ & $-$6.3280(2)$\times10^{-15}$ & $-$5.627(5)$\times10^{-16}$ & $-$4.6220(8)$\times10^{-15}$\\
   Period derivative, $\dot{P}$ & 2.2193(6)$\times10^{-21}$  & 6.0899(2)$\times10^{-20}$ &1.1421(11)$\times10^{-20}$& 2.5407(4)$\times10^{-20}$ \\
   Dispersion measure, DM $\rm(cm^{-3}\rm\,pc)$ & 94.22407(3)$^{\clubsuit}$& 124.7946(8) & 146.732(2)  & 106.4711(6)\\
   Proper motion in $\alpha$, $\mu_{\alpha}$ (mas$\rm\,yr^{-1})$ & $-$7.31(6)& $-$10.0(3)& $-$5.3(8)  &$-$1.7(3)\\
   Proper motion in $\delta$, $\mu_{\delta}$ (mas$\rm\,yr^{-1})$ & 6.76(5)& 2.4(3) & $-$2(3)   & $-$6(3)\\
   Parallax, $\pi$ (mas) & 3.9(12)$^{\diamondsuit}$ & $-$ & $-$ & $-$\\
   \hline
   Orbital period, $P_{\rm{orb}}$ (days) & 6.511905(2)& 76.40321683(5) & 149.1332226(4)  & 0.3111341185(10)      \\ 
   First derivative of orbital frequency, $\dot{n}_{\rm{b}}$ $\rm(Hz\rm\,s^{-1})$  & $-$ & $-$ & $-$ &  1.50(9)$\times10^{-19}$ \\
   Second derivative of orbital frequency, $\ddot{n}_{\rm{b}}$ $\rm(Hz\rm\,s^{-2})$ & $-$ & $-$ & $-$ & $-$5.0(2)$\times10^{-27}$\\
   Projected semi-major axis, $x$ (lt-s) & 4.83004509(11)&33.6383599(8)  & 33.584236(2)    & 0.1201611(6)\\
   Epoch of periastron, $T_{\rm{0}}$ (MJD) & 55335.0641(3)& 55181.5562(15) & 55206.801(10) & 55132.4363(10)\\
   Eccentricity, $e$ & 0.00014204(2)& 0.00025724(3) & 0.00024449(10) & 2.9(6)$\times10^{-5}$\\
   Longitude of periastron, $\omega$ $(\degr)$ &329.682(18)&260.128(7) & 180.00(2)  & 144(12) \\
  Minimum companion mass$^*$, $m_{\rm{c,min}}$ $(M_{\sun})$ &0.19& 0.26 & 0.16  & 0.033\\
  Median companion mass$^{**}$, $m_{\rm{c,med}}$ $(M_{\sun})$ &0.22& 0.31 & 0.19 & 0.039 \\
  Change in $x$, $\dot{x}$ & 9.1(17)$\times10^{-15}$ & $-$ & $-$ & $-$ \\ 
  Variation in $\omega$, $\dot{\omega}$ ($\degr\rm\,yr^{-1}$) & 0.022(9) & $-$&$-$&$-$ \\
\hline
  Binary model & DD & DD & DD & BTX \\
  First TOA (MJD) & 55343.2 & 55131.8 & 55129.1 & 55138.1\\
  Last TOA (MJD) & 56480.0 &  56510.0 & 56491.5 & 56302.1\\
  Timing epoch (MJD) & 55329.1 & 55126.3 & 55132.9 & 55215.1\\
  Points in fit      & 332 & 181     & 99   & 196 \\ 
  Weighted RMS residuals ($\mu$s) &0.8& 5.5 & 7.4 &  3.7\\
  Reduced $\chi^{2}\,^{\ddagger}$ &2.0& 1.5 & 0.7 &  1.9\\ 
\hline
  Mean flux density at 1.4-GHz, $S_{1400}$ (mJy) & 1.00 & 0.86 & 1.31 & 0.37 \\
  Pulse width at 50$\%$ of peak, $W_{\rm{50}}$ ($\degr$) & 10 & 36 & 44& 20\\
\hline
  DM Distance, $d$ (kpc) & 3.0$^{\diamondsuit}$ & 2.6 & 2.8  & 2.5 \\
  Transverse velocity, $V_{\rm{T}}$ $\rm(km\,s^{-1})$ & 140(30) & 120(30) & 70(20)  & 80(40) \\
  Intrinsic period derivative, $\dot{P}_{\rm{int}}$ $(10^{-20})$ & 0.12(2) & 5.94(3) & 0.85(9) & 2.40(7) \\
  Characteristic age$^{\dagger\dagger}$, $\tau_{\rm{c}}$ (Myr)  & 3.1$\times10^{4}$ & 8.3$\times10^{2}$ & 8.4$\times10^{3}$  & 1.5$\times10^{3}$\\
   Spin down energy loss rate$^{\dagger\dagger}$, $\dot{E}$ $(10^{33}\rm\,erg\rm\, s^{-1})$ & 3.7 & 79 & 3.7   & 74\\
   $\dot{E}/d^{2}\,^{\dagger\dagger}$ $(10^{33}\rm\,erg\rm\,kpc^{-2}\rm\,s^{-1})$ & 0.41 & 12 & 0.47  & 12\\
   Characteristic dipole surface magnetic field strength at equator$^{\dagger\dagger}$, $B_{\rm{eq}}$ ($10^{8}$\,G) & 0.53 & 4.3 & 2.0  & 2.4\\
\hline \label{tab:dd} 
\end{tabular}
 \begin{flushleft}
 $^*$ $m_{\rm{c,min}}$ is calculated for an orbital inclination of $i=90^{\circ}$ and an assumed pulsar mass of $1.35\,M_{\sun}$. \\
 $^{**}$ $m_{\rm{c,med}}$ is calculated for an orbital inclination of $i=60^{\circ}$ and an assumed pulsar mass of $1.35\,M_{\sun}$. \\
 $^{\ddagger}$ The reduced $\chi^{2}$ stated here represents the value before the application of EFAC. Note that the rest of the timing solutions have EFACs incorporated, bringing the reduced $\chi^{2}$ to unity. \\
 $^{\clubsuit}$ Temporal DM variations have also been taken into account in the model fit, see explanation in Section~\ref{sec:DM}. \\
 $^{\dagger\dagger}$ These parameters are derived from the intrinsic period derivatives $\dot{P}_{\rm{int}}$. For the derivation of $\dot{P}_{\rm{int}}$ refer to Section~\ref{sec:pdot}.\\
 $^{\diamondsuit}$ We disregard the 3-$\sigma$ $\pi$ measurement when deriving the distance of PSR~J1017$-$7156, as it is likely to be influenced by the Lutz-Kelker bias for example discussed in \citet{Verbiest2010}. \\
 \end{flushleft}
\end{table}
\end{landscape}

\section{UPDATED TIMING OF 12 HTRU MILLISECOND PULSARS} \label{sec:12MSPs}
We have achieved considerable improvement in the timing accuracy for 12 HTRU MSPs compared with results published in their respective discovery papers \citep{HTRU2,HTRU4,Bailes2011}. This is thanks to the now longer timing baseline of more than three years in all cases, and only slightly less for PSR~J1337$-$6423 which has 2.7 years of timing data. The timing parameters resulting from the best fits to the expanded set of TOAs are presented in Tables~\ref{tab:ell1} and \ref{tab:ell2} for pulsars fitted with the ELL1 timing model and in Table~\ref{tab:dd} for pulsars fitted with the DD timing model. 

In the following we discuss the physical implications arising from our timing measurements, including DM variations (Section~\ref{sec:DM}), proper motion and transverse velocities (Section~\ref{sec:PM}), intrinsic period derivatives (Section~\ref{sec:pdot}), binary companions and mass functions (Section~\ref{sec:binary}), Galactic height distributions (Section~\ref{sec:zheight}), orbital eccentricities (Section~\ref{sec:Ecc}), change in projected semi-major axis (Section~\ref{sec:xdot}), orbital period variation (Section~\ref{sec:Pbdot}), variation in the longitude of periastron (Section~\ref{sec:PK}), and gamma-ray associations (Section~\ref{sec:gamma}). 

\subsection{Dispersion measure variations} \label{sec:DM}
Temporal variations in DM, due to turbulence in the ionised interstellar medium (ISM) and the changing line-of-sight to the pulsar, are in theory present in the TOAs of every pulsar \cite[see e.g.,][]{Petroff2013}. However this is typically not observable in slow pulsars since they have limited timing precision. In contrast, for MSPs such variations in DM can become significant and thus require special data treatment \citep{You2007}. 

Indeed for the high-precision timing of PSR~J1017$-$7156 we identified significant temporal variations in its DM measurement, implying changes in the electron density in the ISM along the line-of-sight over a time scale of a few months. We have attempted to model this variation via three correction methods, firstly by fitting DM variations across short ranges of TOAs while holding fixed all other parameters, secondly by including higher order DM derivatives and thirdly by the DM model described in \citet{Keith2013}. In Figure~\ref{fig:J1017-DM} we plot the manually identified values of DM across every few TOAs in black. We plot the best-fit curve from the timing solution of \textsc{tempo2}, employing up to eight DM-derivatives as the green dashed line. We plot the DM model derived using the method outlined in \citet{Keith2013} as red crosses, and the red solid line joining them shows the resulting DM model. It can be seen that the DM derivatives provide a smooth fit to the DM variations, however there are still small scale variations that are not properly accounted for. On the other hand, the DM model essentially creates a linear interpolation between DM offsets identified at specific epochs (note that here we have adopted a gap of 50\,days between successive DM offsets), and hence can be tailor-made to follow more closely variations on all scales. We conclude that the DM model of \citet{Keith2013} gives a more successful fit and hence have adopted this for the timing solution of PSR~J1017$-$7156.    

\begin{figure}
\includegraphics[width=3.3in]{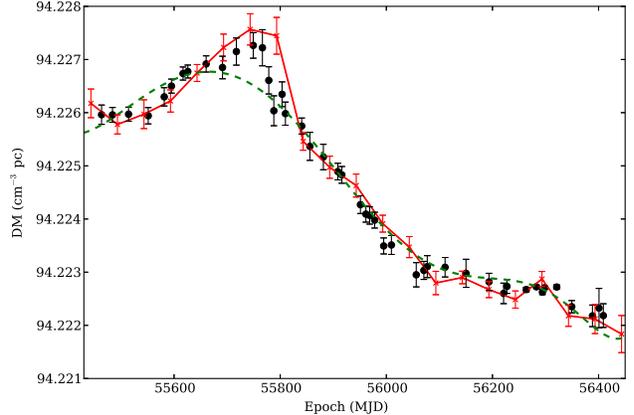} 
\caption{DM for J1017$-$7156 with time. The manually identified DM variations across every few TOAs are plotted as black filled circles. The green dashed line shows the best-fit curve from timing solution generated in \textsc{tempo2}, employing DM-derivative terms up to the eighth order. The red crosses are the DM offsets identified by applying the method in accordance with the description in \protect\citet{Keith2013} and the red solid line joining them shows the resulting DM model.}  
\label{fig:J1017-DM}
\end{figure}

\subsection{Proper motion and transverse velocities} \label{sec:PM}
The proper motion $(\mu)$ of a pulsar introduces a positional offset over time and is measurable from pulsar timing data. Within our sample of 12 MSPs with extended timing solutions, we have measured five new proper motions with significances greater than 3-$\sigma$, for PSRs~J1017$-$7156, J1125$-$5825, J1446$-$4701, J1708$-$3506 and J1719$-$1438. PSR~J1811$-$2405 is very close to the ecliptic plane with $(\lambda,\beta)=(272.586\degr, -0.675\degr)$ which means its proper motion in ecliptic latitude ($\mu_{\beta}$) cannot be well-constrained. With a $\lambda$ so close to 270$\degr$, the translation from ecliptic frame to equatorial frame would have almost no rotation. This implies that the large uncertainty associated with $\beta$ is only inherited in the declination, $\delta$, without also contaminating the right ascension, $\alpha$. Hence for PSR~J1811$-$2405 we can choose to continue using the equatorial coordinates and we fixed $\mu_{\delta}$ at zero for the rest of the analysis. For the four newly-discovered MSPs in this paper, their time spans are not yet long enough for proper motion to be detected with significance.

From $\mu$ and their respective pulsar distances, $d$, we can derive their corresponding transverse velocities, $V_{\rm{T}}$, with the following equation, 
\begin{equation}
V_{\rm{T}} = \rm4.74\,km\,s^{-1} \times \left(\frac{\mu }{\rm mas\,yr^{-1}}\right) \times \left(\frac{d}{\rm kpc}\right)\,. \label{eq:Vt} 
\end{equation}
In this work we have calculated pulsar distances based on the NE2001 electron density model \citep{NE2001model} and we assume an associated uncertainty of 25$\%$ for each DM-derived distance. MSP proper motion measurements are relatively rare and hence there are not many derived velocities (only about 40 currently published values in the literature), making it difficult to place constraints on MSP velocity distribution models. The latest MSP velocity discussions can be found in \citet{Toscano1999} and \citet{Hobbs2005}, proposing an average velocity for recycled MSPs of $85\pm13\rm\,km\,s^{-1}$ and $87\pm13\rm\,km\,s^{-1}$ respectively. \citet{Hobbs2005} also quoted a median velocity for recycled MSPs of $73\rm\,km\,s^{-1}$. Our new $V_{\rm{T}}$ measurements largely agree with these previous results (refer to Table~\ref{tab:ell1}~to~\ref{tab:dd}). Note that we believe the high V$_{\rm{T}}$ of 670 and 350$\rm\,km\,s^{-1}$ reported for PSRs~J1708$-$3506 and J1731$-$1847 in \citet{HTRU2} should in fact be corrected to more modest values of $70\pm20$ and $80\pm40\rm\,km\,s^{-1}$ respectively. 

\subsection{Observed and inferred intrinsic period derivatives} \label{sec:pdot}
The vast majority of pulsars are rotation-powered objects and hence their respective period derivatives $(\dot{P})$ are fundamental to their identities. The observed period derivatives $(\dot{P}_{\rm{obs}})$ however contain a contribution from kinematic effects \citep{Shk1970} and acceleration due to the Galactic potential \citep{DT1991}. Determination of the intrinsic period derivative is important for properly placing pulsars in the $P$-$\dot{P}$~diagram from which physical conclusions (such as magnetic field strength, characteristic ages) may be drawn. To obtain the intrinsic period derivative $(\dot{P}_{\rm{int}})$ we employed the following equation,
\begin{equation}
\dot{P}_{\rm{int}} = \dot{P}_{\rm{obs}} -  \dot{P}_{\rm{shk}} -  \dot{P}_{\rm{gal}} \label{eq:PdotObs}\,.   
\end{equation}
The term $\dot{P}_{\rm{shk}}$ accounts for the apparent acceleration that arises from the transverse motion of the pulsar. It is related to the pulsar spin period, $P$, the proper motion, $\mu$, and the pulsar distance, $d$, by the following equation from \citet{Shk1970},  
\begin{equation}
\dot{P}_{\rm{shk}} =  \left( \frac{P}{c} \right) d\,\mu^2\,.       
\end{equation}
The term $\dot{P}_{\rm{gal}}$ accounts for difference in the line-of-sight components of the acceleration of the pulsar and the Solar System under the influence of the Galactic gravitational potential. There exist several Galactic potential models in the literature, and we have chosen the one described in \citet{Paczynski1990}. This model reproduces a flat rotation curve and uses a Solar Galactocentric distance $R_{0}$ of 8\,kpc and a Solar Galactic rotation velocity of $220\rm\,km\,s^{-1}$.  

\begin{table}
\setlength{\tabcolsep}{0.15cm}
\centering
 \caption{Table listing the derived $\dot{P}_{\rm{shk}}$ and $\dot{P}_{\rm{gal}}$ for the 12 MSPs with updated timing solutions. The final column shows the inferred $\dot{P}_{\rm{int}}$. Values in parentheses are the nominal 1-$\sigma$ uncertainties in the last digits.} 
\begin{tabular}{p{1.4cm}R{0.4cm}@{}p{0.9cm}R{0.6cm}@{}p{1.2cm}R{0.4cm}@{}p{0.7cm}R{0.7cm}@{}p{0.9cm}}
  \hline
  PSR && $\dot{P}_{\rm{obs}}$ & & $\dot{P}_{\rm{shk}}$ & & $\dot{P}_{\rm{gal}}$ & & $\dot{P}_{\rm{int}}$ \\
      && ($\rm10^{-20}$) & & ($\rm10^{-20}$) & & ($\rm10^{-20}$) & & ($\rm10^{-20}$) \\ 
  \hline
J1017$-$7156 & 0 &.22193(6)   & 0 &.16(4)  & $-$0 & .067(15) & 0 &.12(2)  \\
J1125$-$5825 & 6 & .0899(2)   & 0 &.20(5)  & $-$0 & .066(19) & 5 &.94(3) \\
J1337$-$6423 & 2 & .47(2)     & 1&.8(13)   & $-$0 & .40(15)  & 1 &.0(13) \\
J1446$-$4701 & 0 & .9810(2)   & 0 &.016(4) & $-$0&.007(2)    & 0 &.972(2)  \\
J1502$-$6752 & 31&.45(13)     & 16&(14)    & $-$0&.8(3)      & 15&(14)   \\
J1543$-$5149 & 1 & .6161(14)  & 0&.06(3)   & 0&.0124(13)     & 1&.54(3) \\ 
J1622$-$6617 & 5 & .88(2)     & 1&.0(8)    & $-$0&.17(7)     & 5&.0(8) \\
J1708$-$3506 & 1 & .1421(11)  & 0&.14(7)   & 0 &.14(4)       & 0 &.85(9)      \\
J1719$-$1438 & 0 & .8044(8)   & 0&.23(10)  & 0 &.028(9)      & 0 &.54(11) \\
J1731$-$1847 & 2 &.5407(4)    & 0&.08(6)   & 0 &.048(14)     & 2 &.40(7)\\
J1801$-$3210 & $-$0&.004(4)   & 2&.3(16)   & 0 &.41(15)      & $-$2&.7(17)$^{*}$ \\
J1811$-$2405 & 1 &.33780(16)  & 0&.00035(18)$^{\dagger}$& 0&.052(15) & 1&.284(15) \\
\hline \label{tab:pdot}
 \end{tabular}
\vspace{-0.8\skip\footins}
 \begin{flushleft}
 $^{*}$ The potential causes of this apparent negative period derivative are discussed in the main text of Section~\ref{sec:pdot1801}.\\
 $^{\dagger}$ This is a lower limit of $\dot{P}_{\rm{shk}}$ since PSR~J1811$-$2405 is very close to the ecliptic plane (refer to Section~\ref{sec:PM}). Its $\mu_{\delta}$ cannot be constrained and is fixed to zero.  \\
 \end{flushleft}
\end{table}

\begin{figure}
\includegraphics[width=3.3in]{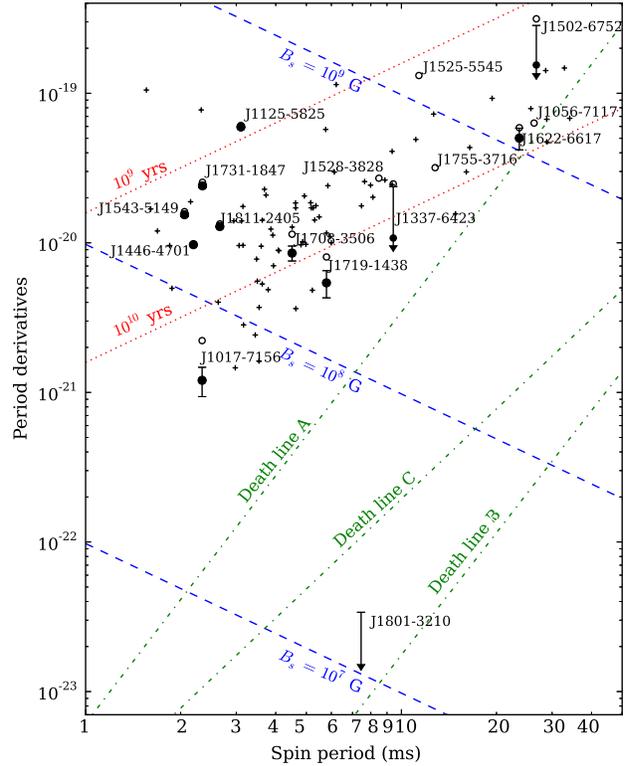}
\caption{The $P$-$\dot{P}$~diagram plotted for the region of MSPs. Black open circles show the $\dot{P}_{\rm{obs}}$ for all 16 MSPs in this work, except PSR~J1801$-$3210 for which a 2-$\sigma$ upper limit is shown because we have a measured $\dot{P}_{\rm{obs}}$ value consistent with zero within 1-$\sigma$ even with 4 years of timing data. For the 12 MSPs in this work with updated timing solutions, we are able to plot also their corrected locations of $\dot{P}_{\rm{int}}$ in the $P$-$\dot{P}$~diagram represented by black filled circles with associated error bars. Two of the MSPs (PSRs~J1337$-$6423 and J1502$-$6752) have unconstrained $\dot{P}_{\rm{int}}$, hence we plot the 95\% confidence upper limit. Note that PSR~J1801$-$3210 has an apparent negative $\dot{P}_{\rm{int}}$ even at the 95\% confidence upper limit therefore we show only its $\dot{P}_{\rm{obs}}$. The red dotted lines correspond to characteristic ages of $10^{9}$ and $10^{10}$\,years respectively, whereas the blue dashed lines show derived surface magnetic field strength at the equator $(B_{\rm{eq}})$ of $10^{7}$, $10^{8}$ and $10^{9}$\,G. Both of these sets of lines are derived according to equations in \protect\citet{Handbook2004}. The green dot-dashed lines plot the three pulsar death lines as described in \protect\citet{Chen1993}, derived from the theoretical relationship between surface magnetic field strength at the polar region $(B_{\rm{p}})$ and pulsar spin period ($P$).}
\label{fig:ppdot}
\end{figure}

Table~\ref{tab:pdot} lists the $\dot{P}$ contributions as calculated for the 12 MSPs with updated timing solutions in our sample. Monte Carlo simulations with 1,000,000 runs per pulsar have been used to estimate the associated error. Note that the errors in $\dot{P}_{\rm{shk}}$ and $\dot{P}_{\rm{gal}}$ do not reflect the effect of errors in the distance estimates. The results are illustrated in Figure~\ref{fig:ppdot}, which is a $P$-$\dot{P}$~diagram around the region where MSPs are located. The $\dot{P}_{\rm{obs}}$ and the corrected $\dot{P}_{\rm{int}}$ of the 12 MSPs studied in this paper are plotted, together with other known pulsars in this region. 

Some of the results (noticeably those of PSR~J1337$-$6423 and J1502$-$6752) have large associated errors and should be considered with caution. One reason is that $\dot{P}_{\rm{shk}}$ relies on the square of $V_{\rm{T}}$, which is in turn dependent on proper motion as seen from Equation~(\ref{eq:Vt}). Hence $\dot{P}_{\rm{shk}}$ is only meaningful for MSPs with well-constrained proper motion measurements. Additionally, $\dot{P}_{\rm{gal}}$ is dependent on the distance of the pulsar, $d$. As mentioned in Section~\ref{sec:PM}, the DM-derived distance is thought to have $\sim25\%$ error, and can be much larger for individual pulsars. 

\subsubsection{PSR~J1017$-$7156} \label{sec:pdot1017}
Disregarding these two unconstrained measurements, PSR~J1017$-$7156 stands out with one of the smallest inferred intrinsic $\dot{P}$ at a value of $1.2\times10^{-21}$. We are aware that if red noise is present in the data this could potentially also contaminate our $\dot{P}$ measurement. However if we include the frequency second derivative in the model fit in an attempt to whiten the data with a quadratic component, the $\dot{P}$ measurement remains statistically consistent. PSR~J1017$-$7156 is thus located at the bottom left of the $P$-$\dot{P}$~diagram, which yields a characteristic age, $\tau_{\rm{c}} \equiv$ $P/(2\dot{P})$, of 31\,Gyr, i.e. larger than the Hubble age. Note that $\tau_{\rm{c}}$ is by no means a reliable age indicator for MSPs, since it is only applicable for pulsars which have a braking index $n=3$ and an initial spin period ($P_{0}$) much less than the current spin period, which is not thought to be the case for MSPs. However for MSPs with such small $\dot{P}_{\rm{obs}}$ like that of PSR~J1017$-$7156, we can deduce that the MSP was probably born with small initial period derivative and must not have moved very far from its current location on the $P$-$\dot{P}$~diagram since its birth \citep{Tauris2012}. The derived surface magnetic field strength at the equator\footnote{Note throughout the paper we differentiate between the derived surface magnetic field at the polar region ($B_{\rm{p}}$), and that of the equatorial region ($B_{\rm{eq}}$) which is only half the strength comparing to the polar region.} of PSR~J1017$-$7156 is also at one of the lowest known at $5.3\times10^{7}$ G.

\subsubsection{PSR~J1801$-$3210} \label{sec:pdot1801}
There is one peculiar case, PSR~J1801$-$3210, for which no significant period derivative has been measured, even with more than 4 years of timing data. The best-fit solution in \textsc{tempo2} shows a $\dot{P}_{\rm{obs}}$ of $-4\pm4\times10^{-23}$, an extremely small number compared to that of typical MSPs ($\dot{P}_{\rm{obs}}$ of the order of $10^{-19}$ to $10^{-20}$). A 2.7-$\sigma$ $\dot{P}$ value of $0.265(97)\times10^{-20}$ was presented in the initial discovery paper by \citet{HTRU2} which at that time had just over one year of timing data, however this value is inconsistent with our current longer time baseline TOAs. Shortening our data span to the same epoch as that in \citet{HTRU2} results in an unconstrained $\dot{P}_{\rm{obs}}$ measurement of 0.2(20)$\times10^{-20}$, eliminating the possibility of an actual change in period derivative over time.

Referring again to Equation~(\ref{eq:PdotObs}), proper-motion-induced $\dot{P}_{\rm{shk}}$ has an always positive contribution to $\dot{P}_{\rm{obs}}$, so that $\dot{P}_{\rm{int}}$ will be even smaller. $\dot{P}_{\rm{gal}}$ however could have a positive or negative contribution depending on the relative location of the pulsar in the Galaxy with respect to the Earth. 

PSR~J1801$-$3210 has a proper motion measurement of $15(7)\rm\,mas\rm\,yr^{-1}$, corresponding to a positive $\dot{P}_{\rm{shk}}$ of the order of $10^{-20}$. The \citet{Paczynski1990} Galactic potential model shows that at the NE2001 DM-derived distance of 4\,kpc PSR~J1801$-$3210 would be accelerated away from the Sun, giving a positive $\dot{P}_{\rm{gal}}$ of the order of $10^{-21}$ (Table~\ref{tab:pdot}) to further decrease the already negative $\dot{P}_{\rm{obs}}$. Even if we assume the proper motion to be zero to get the smallest possible contribution from $\dot{P}_{\rm{shk}}$, we still cannot overcome this apparent negative $\dot{P}_{\rm{int}}$ at the given DM distance of 4\,kpc, since the $\dot{P}_{\rm{gal}}$ is positive and dominates the tiny $\dot{P}_{\rm{obs}}$ of $10^{-23}$. We acknowledge that the \citet{Paczynski1990} model consists basically of only three elements: a bulge, a disk and the surrounding halo. However this is considered a valid approximation, and for example the effect of spiral arm structure should not significantly skew the model. 

In the following we consider other potential explanations to this apparent negative $\dot{P}_{\rm{int}}$, i.e. effects that would have contributed to the $\dot{P}_{\rm{obs}}$ but are not yet accounted for in Equation~(\ref{eq:PdotObs}). We discuss the cases of (a) acceleration due to local stars; (b) acceleration due to giant molecular clouds (GMCs); and (c) acceleration due to a third orbiting object if PSR~J1801$-$3210 is in a triple system. 
 
If there exists a third body (with mass $M_{3}$) located near the pulsar, in a direction towards the Earth and close to the line-of-sight, it will potentially cause a radial acceleration of PSR~J1801$-$3210 towards the Earth. We can express the mass of the third body required to produce a $\dot{P}$ contribution of $\dot{P}_{\rm{M_{3}}}$ as,
\begin{equation}
M_{3} = \left( \frac{\dot{P}_{\rm{M_{3}}}}  {P_{\rm{spin}}} \right) \left( \frac{c\,r^{2}} {G} \right) (\cos\theta)^{-1}\,,  \label{eq:M3}
\end{equation}
where $r$ is the distance between the third body and the pulsar, $G$ is Newton's gravitational constant and $\theta$ is the angle between the direction from the pulsar to the third body and the direction from the pulsar to the Sun. We imagine the scenario of $\theta \approx 0\degr$ where the line-of-sight acceleration induced on the pulsar is the largest, and we first examine the potential contribution from stars located near the pulsar. The probability distribution of fluctuation in Galactic acceleration due to local clustering centres has been studied in the literature \citep[see e.g.,][]{Holtsmark1919}, and based on Equation~(3.1) and (3.5b) in \citet{DT1991} one finds for PSR~J1801$-$3210 at 1-$\sigma$ confidence level,
\begin{equation}
 \abs*{\dot{P}_{*}}_{1-\sigma} = 3.3\times10^{-24}   \left( \frac{\hat{M}}{M_{\sun}} \right)^{1/3} \left(\frac{\rho}{\rho_{\sun}} \right)^{2/3}\,, \label{eq:starnearby}   
\end{equation}
where $\dot{P}_{*}$ is the potential period derivative contribution from nearby stars, $\hat{M}$ is the average of mass taken over the mass spectrum of the attracting centres and we use the same value of 1\,$M_{\sun}$ as in \citet{DT1991}. The local stellar-mass density, $\rho_{\sun}$, has a value of 0.06$\,M_{\sun}\,\rm{pc}^{-3}$ according to \citet{Mihalas1981} and the stellar-mass density, $\rho$, can be extrapolated by,
\begin{equation}
\rho =\rho_{\sun} \times \exp \left( \frac{R_{0} - R}{L_{\rm{disk}}} \right) \exp \left(- \frac{z}{z_{\rm{h}}} \right)\,,\label{eq:rho}
\end{equation}
where $R_{0}$ is the aforementioned Solar Galactocentric distance at 8\,kpc. $L_{\rm{disk}}$ is the stellar disk scale length and $z_{\rm{h}}$ is the scale height of the stellar disk component, which from the most recent literature by \citet{Bovy2013} $L_{\rm{disk}} = 2.15\rm\,{kpc}$ and $z_{\rm{h}} = 0.4\rm\,{kpc}$. $R$ is the distance of the pulsar from the Galactic Centre, and for the case of PSR~J1801$-$3210 it is approximately 4\,kpc as derived from the NE2001 model. This corresponds to a Galactic height, $z$, of 0.32\,kpc. Substituting these into Equations~(\ref{eq:rho}) and (\ref{eq:starnearby}) gives $\rho = 0.17\,M_{\sun}\,\rm{pc}^{-3}$ and a tiny $\dot{P}_{*}$ of the order of $10^{-24}$ which is unlikely to have led to the negative $\dot{P}_{\rm{int}}$. To appreciate the improbability of this scenario we can also hypothesise a nominal $\dot{P}_{*}$ of the order of $-10^{-21}$. From Equation~(\ref{eq:starnearby}) this would require $\rho$ to be more than 300$\,M_{\sun}\,\rm{pc}^{-3}$ and nowhere along the line-of-sight direction of PSR~J1801$-$3210 has such high stellar-mass density.

Alternatively let us consider the contribution from GMCs, and again we assume that there exists such an acceleration acting upon the pulsar towards the Earth which induces a nominal $\dot{P}_{\rm{GMC}}$ of the order of $-10^{-21}$. GMCs typically have masses between $10^{3}$ to $10^{7}\,M_{\sun}$; substituting this into Equation~(\ref{eq:M3}) corresponds to a distance, $r$, of about 2 to 190\,pc from the pulsar. No GMCs are known to exist near PSR~J1801$-$3210, but not all GMCs have necessarily been detected, so this possibility cannot be ruled out. It may also be that multiple smaller molecular clouds (also known as Bok globules) act together to accelerate PSR~J1801$-$3210 in our direction. 

Another possible candidate of this third body could be a tertiary star or an exoplanet orbiting PSR~J1801$-$3210 in a weakly-bounded hierarchical triple orbit. This third component would accelerate the pulsar system towards it, and hence if the third component happened to provide a net acceleration on PSR~J1801$-$3210 towards the Earth it would lead to the negative $\dot{P}_{\rm{int}}$ like in the case of a GMC as mentioned above. We can achieve the same $\dot{P}_{\rm{exo}}$ of the order of $-10^{-21}$, for example with an Earth-sized exoplanet at distance of $\sim$20\,AU in an orbit of $\sim$70\,years around PSR~J1801$-$3210, or a Jupiter-sized exoplanet at a distance of $\sim$400\,AU in a large orbit of $\sim$6000\,years, assuming circular orbit. 

The relative motion between the pulsar system and the exoplanet would have induced variations in the acceleration, as well as variations in the second derivative of spin frequency, $\ddot{\nu}$ \citep{Backer1993}. We do not have a significant measurement of $\ddot{\nu}$ except a 2-$\sigma$ upper limit of $8\times10^{-26}\rm\,s^{-3}$. This thus excludes the existence of a nearby exoplanet and favours the case of a further-out heavier object. However at the same time, for a third orbiting object to stay bound with the pulsar system, a very strict limit on the post-supernova (SN) recoil velocity of the inner binary is required \citep{Hills1983}. Precisely, the recoil velocity has to be no more than $30\rm\,km\rm\,s^{-1}$ and 7$\rm\,km\rm\,s^{-1}$ for the case of an Earth-mass and a Jupiter-mass exoplanet respectively. According to simulations by \citet{Tauris1996}, the recoil velocity of any surviving binary is expected to be larger than 20$\rm\,km\rm\,s^{-1}$, even for a symmetric SN explosion, unless the pulsar formed via an accretion-induced collapse of a white dwarf \citep{Nomoto1979}. Hence, we are inclined to exclude a very distant third body with a Jupiter mass, and notice that a closer Earth-mass object would require quite some fine-tuning in the SN event to remain bound. To summarise, we conclude that this scenario of an exoplanet is possible but unlikely.

\begin{figure}
\includegraphics[width=3.3in]{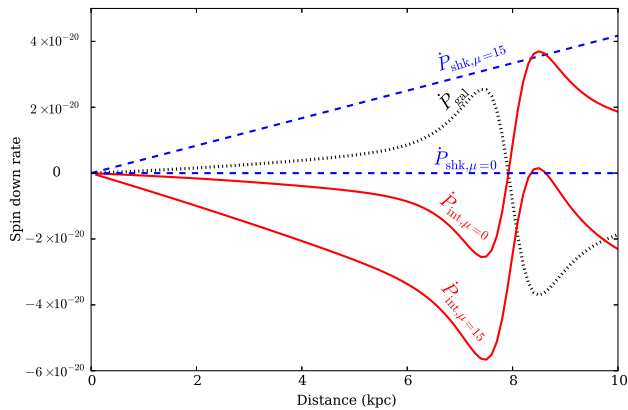} 
\caption{Plot showing various $\dot{P}$ contributions for PSR~J1801$-$3210. The black dotted line is the $\dot{P}_{\rm{gal}}$ as a function of distance and is independent of proper motion. The two blue dashed lines show the $\dot{P}_{\rm{shk}}$ caused by a proper motion $(\mu)$ of 0 and 15$\rm\,mas\rm\,yr^{-1}$ respectively. The two red solid lines show the resulting $\dot{P}_{\rm{int}}$. In the case of $\mu=15\rm\,mas\rm\,yr^{-1}$ the corresponding $\dot{P}_{\rm{int,\mu=15}}$ is always negative. In the case of no proper motion ($\mu=0\rm\,mas\rm\,yr^{-1}$) the corresponding $\dot{P}_{\rm{int,\mu=0}}$ can become positive only after a distance of at least 8\,kpc.}
\label{fig:ppdot-J1801}
\end{figure}

Finally, we consider the possibility that the NE2001 DM-derived distance of 4\,kpc is significantly wrong, hence locating PSR~J1801$-$3210 in a different quadrant of the Galaxy which would reverse the direction of the Galactic potential and the sign of $\dot{P}_{\rm{gal}}$. In Figure~\ref{fig:ppdot-J1801} we plot the various $\dot{P}$ contribution as a function of distance along the line-of-sight of PSR~J1801$-$3210. It can be seen that in the limiting case of $\dot{P}_{\rm{shk}}$ being zero, we can achieve a positive period derivative beyond a distance of 8\,kpc, and can reach an upper limit of $\dot{P}_{\rm{int}}$ of $3\times10^{-20}$ at a distance of 8.5\,kpc. At a distance of 8\,kpc, the NE2001 model requires a corresponding DM of 326.1$\rm\,cm^{-3}\rm\,pc$ which is inconsistent with the well-constrained DM measurement of PSR~J1801$-$3210 of only 177.713(4)$\rm\,cm^{-3}\rm\,pc$. However other electron density models give very different results. For example the TC93 model \citep{TC93} requires a corresponding DM of only 227.0$\rm\,cm^{-3}\rm\,pc$, whereas including a thick disk component to the TC93 model \citep{Schnitzeler2012} predicts an even smaller corresponding DM of 185.5$\rm\,cm^{-3}\rm\,pc$, which is only a factor of 1.07 from our measured value. These large discrepancies between various models reflect uncertainties in the electron density distribution along this line-of-sight, and thus it seems plausible that the DM-derived distances of PSR~J1801$-$3210 have been underestimated. PSR~J1801$-$3210 is located at $(l, b)=(358.922\degr, -4.577\degr)$, a distance of at least 8\,kpc in this direction would put PSR~J1801$-$3210 just beyond the Galactic Centre, hence reversing the direction of $\dot{P}_{\rm{gal}}$. In any case, we suggest that PSR~J1801$-$3210 would serve as an important test pulsar for improving future electron density models.

Otherwise, if PSR~J1801$-$3210 has indeed an extremely small $\dot{P}_{\rm{int}}$ it would imply an exceptionally small surface magnetic field. Popular theories on the pulsar emission mechanism require electron-positron pair production, and the longer the spin period of the pulsar, the larger the potential needed to power the particle acceleration \citep[see for example][]{Beskin1988}. The following implication is known as the `pulsar death line', which predicts for a particular pulsar spin period, there exists a lower limit of period derivative and surface magnetic field for which radio emission can be produced. Therefore, we can derive a lower limit of $\dot{P}_{\rm{int}}$ for PSR~J1801$-$3210 to stay above the pulsar death line. We adopt the theoretical study from \citet{Chen1993} which described three possible death lines also plotted in Figure~\ref{fig:ppdot}. If we take the lowest limiting case imposed by death line B, we derive a lower limit of $\dot{P}_{\rm{int}}=7.9\times10^{-24}$ and a corresponding surface magnetic field at the equator ($B_{\rm{eq}}$) of $7.8\times10^{6}$\,G. We note that this derivation assumes a contribution only from a model of a vacuum magnetic dipole. However as discussed by \citet{Tauris2012}, if the spin-down torque caused by the plasma current in the magnetosphere \citep{Spitkovsky2006} is also taken into account, the realistic surface magnetic field would even be lower, by at least a factor of $\sqrt{3}$.  

\subsection{Binary companions and mass functions } \label{sec:binary} 
\begin{figure*}
\centering
\setlength\fboxsep{0pt}
\setlength\fboxrule{0pt}
\fbox{\includegraphics[width=13cm]{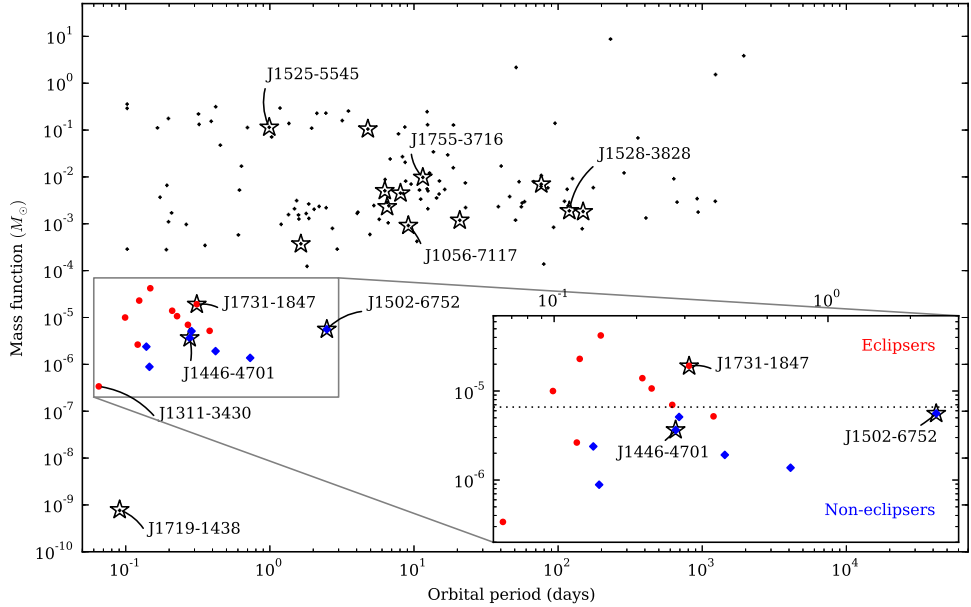}}
\caption{Plot showing mass function vs orbital period for all binary pulsars. Known `eclipsers' are represented by red circles and `non-eclipsers' by blue diamonds. The 16 MSPs studied in this paper are represented by star symbols. The zoomed-in panel focuses on the region of the VLMBPs, and the dotted line plotted within represents the dividing mass function value of $6.7\times10^{-6}\,M_{\sun}$, which corresponds to $m_{\rm{c}}=0.029\, M_{\sun}$ and $M_{\rm{p}}=1.7\,M_{\sun}$ assuming an orbital inclination of 70$\degr$.}
  \label{fig:mfpb}
\end{figure*}

A plot of mass function versus orbital period is a standard way of distinguishing different types of binary systems and can be used to gain insight into the nature of the binary companion, as shown in Figure~\ref{fig:mfpb}. Indeed it can be seen immediately that PSR~J1719$-$1438 occupies an otherwise empty region in the bottom left corner of this figure, as a result of its uniquely light, planet-mass companion. This has been extensively discussed in the literature \citep[e.g.,][]{Bailes2011,vHaaften2012} so will not be further elaborated in this paper.

A cluster of pulsars can be seen in the left side of Figure~\ref{fig:mfpb}, with $P_{\rm{orb}}$ $\le$ 1\,day and mass functions between $10^{-7}$ to $10^{-4}\,M_{\sun}$. They are considered descendants of close LMXB systems, resulting in the formation of a binary with an ultra-light companion \citep{Tauris2011}, also known as the `very low-mass binary pulsars' (VLMBPs). In our sample we have three MSPs that fit into this category, namely PSRs~J1446$-$4701, J1502$-$6752 and J1731$-$1847.    

Some of the VLMBPs exhibit eclipses and are typically referred to as black widow pulsars \citep[BW;][]{Roberts2013}. Eclipses have already been reported for PSR~J1731$-$1847 by \citet{HTRU2}, but not for PSRs~J1446$-$4701 nor J1502$-$6752. \citet{Freire2005} proposed a correlation between the possibility of observing eclipses and orbital inclination for these VLMBPs in GC. The essence of the idea is that the companions of these VLMBPs have a narrow intrinsic mass distribution, and subsequently whether a VLMBP shows eclipses or not, becomes exclusively dependent on its orbital inclination. In other words, a VLMBP viewed relatively face-on (low inclination) is less likely to be observed as an eclipsing system and will also have a smaller mass function, and vice versa. 

While this hypothesis seems to work well in GCs, there has not yet been a similar study on the non-GC associated VLMBP population. We have compiled all related literature, and colour-coded in Figure~\ref{fig:mfpb} the known `eclipsers' as red circles and the `non-eclipsers' as blue diamonds. Two distinct groups composed of `eclipsers' and `non-eclipsers' do seem to exist, with only one outlier, PSR~J1311$-$3430, which is the tightest binary pulsar known with a $P_{\rm{orb}}$ of just 93\,minutes \citep{Romani2012,Pletsch2012,Ray2013}. But this pulsar may have evolved from an ultra-compact X-ray binary (UCXB), hence belonging to a different population \citep{vanHaaften2012} and might not be applicable to the hypothesis as mentioned above. Disregarding this system, it is striking to see a bimodal distribution. Particularly interesting is that there is no non-eclipsing system found within the red cluster of `eclipsers', although from a pulsar searching point-of-view these kinds of systems should in fact be easier to detect due to their non-eclipsing nature.

Plotted as a dotted line in the zoomed-in panel of Figure~\ref{fig:mfpb} is our nominal split between the `eclipsers' and the `non-eclipsers', representing a dividing mass function of $6.7\times10^{-6}\,M_{\sun}$. We assume a pulsar mass of $1.7\,M_{\sun}$, and an orbital inclination of 70$\degr$ to postulate a lower limit on inclination which eclipses can be observed. This dividing mass function would then correspond to $m_{\rm{c}}= 0.029\,M_{\sun}$, which is also within the range of typical companion masses of BWs as shown in \citet{Chen2013}. Indeed orbital eclipses are observed for PSR~J1731$-$1847 which has a median $m_{\rm{c}}$ of 0.0385\,$M_{\sun}$ and lies above the dotted line, whereas no eclipse is observed for PSRs~J1446$-$4701 and J1502$-$6752 with lower companion masses (median $m_{\rm{c}}$ of 0.022\,$M_{\sun}$ and 0.025\,$M_{\sun}$ respectively) located below this line. These measurements are in agreement with \citet{Freire2005}. 

\subsection{Galactic height distribution} \label{sec:zheight}
Based on theoretical grounds we expect an anti-correlation between the absolute Galactic height and the inferred mass function of binary pulsars. The reason is the following: assuming that the momentum kick imparted to a newborn neutron star during the SN explosion is independent of exterior parameters, such as the mass of the companion star, the resulting systemic recoil velocity is larger for systems with smaller companion star masses (and thus smaller mass functions) as a simple consequence of conservation of momentum. Since the acquired amplitude of the Galactic motion of the system only depends on the systemic recoil velocity, we therefore expect the above mentioned anti-correlation between the distribution of observed Galactic heights and the measured mass functions of pulsar binaries. Some theoretical studies \citep[e.g.,][]{Tauris1996} have suggested the possibility of a weak relation between orbital period and systemic recoil velocity of pulsar binaries. However, \citet{Gonzalez2011} found no observational evidence for such a relation based on the 2D velocities of binary MSPs. Thus we disregard orbital periods in the following discussion.

\begin{figure*}
  \centering
  \setlength\fboxsep{0pt}
  \setlength\fboxrule{0pt}
  \fbox{\includegraphics[width=17cm]{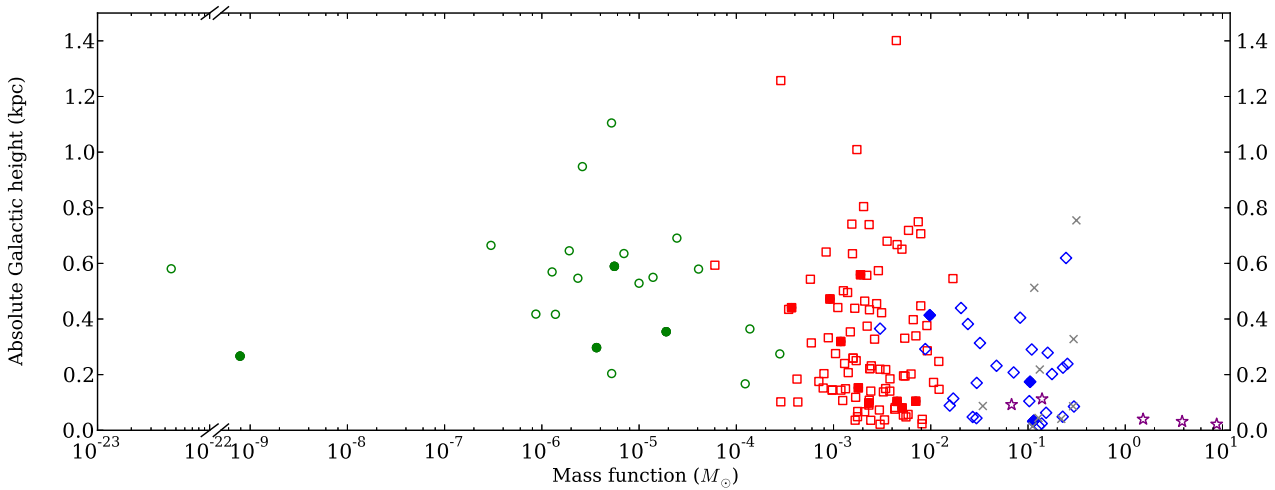}}
\caption{Mass function vs absolute Galactic height from the plane, |$d\sin{b}$|. We derived the distances, $d$, according to the \protect\citet{NE2001model} NE2001 model of the Galactic electron density, except for 19 binary systems for which independent distance measurements existed. In those cases we used the independently-measured distances instead of the DM-derived distances. Ultra-light systems are plotted as green circles, binaries with He-WD companions as red squares, massive CO or ONeMg-WD companions as blue diamonds, and main-sequence star companions as purple stars. The 16 MSPs in this work are also plotted with the same scheme, but emphasised by filling the symbols with the relevant colours. NS-NS systems are plotted as grey crosses but since they have received two kicks from SN explosions they are not considered further in this discussion.}
\label{fig:mfgb}
\end{figure*}

Our sample of 16 MSPs has a wide distribution of mass functions, from PSR~J1719$-$1438 with an ultra-low mass companion and a mass function of 7.8$\times10^{-10}\,M_{\sun}$ to PSR~J1525$-$5545 with a massive WD companion and a mass function of 0.11$\,M_{\sun}$. With the addition of these systems, we investigate whether there exists a correlation between mass function and vertical distance from the Galactic plane $(|d\sin{b}|)$. We have taken our sample of MSPs from the \textit{ATNF Pulsar Catalogue}\footnote{http://www.atnf.csiro.au/people/pulsar/psrcat/} \citep{PSRCAT} and an online MSP catalogue maintained by Lorimer\footnote{http://astro.phys.wvu.edu/GalacticMSPs/GalacticMSPs.txt}. We have included the 16 MSPs in this work and also six additional newly-discovered HTRU MSPs \citep{Ng,Thornton}. All recycled MSPs in binary systems are considered, provided that they are not associated with a GC or extragalactic, which amounts to 164 MSPs in total. We continue to use the \citet{NE2001model} model of Galactic electron density to derive the distances of all known pulsars in order to calculate their respective Galactic heights. Independent distance measurements are available for 19 binary systems and we use these, instead of the DM distances, when calculating their Galactic heights. In Figure~\ref{fig:mfgb} we plot the absolute Galactic heights against mass functions, and we classify the nature of each of the binary companions in accordance with the description in \citet{Tauris2012}. This results in five binary groups, namely those with ultra-light (UL) companions, with He-WD companions, with massive CO or ONeMg-WD companions, neutron-star$-$neutron-star (NS-NS) systems and those with main-sequence star (MS) companions. For the rest of the discussion we set aside the nine NS-NS systems, since they were born with two SN explosions (hence received two kicks) and would complicate our discussion.

\begin{table}                                                                                 
\centering                                                      
 \caption{A summary of the statistical distribution of Galactic height for each binary group, classified in accordance with the description in \protect\citet{Tauris2012}. $N$ is the number of pulsar systems in each group. The average (|$z_{\rm{mean}}$|) and the median (|$z_{\rm{med}}$|) Galactic heights in kpc are listed, as well as the corresponding standard deviation ($\sigma$).}
 \begin{tabular}{llp{0.9cm}p{0.9cm}l}                                        
  \hline                                                                                       
 Binary group & N & $z_{\rm{mean}}$ (kpc) & $z_{\rm{med}}$ (kpc) & $\sigma$ \\                                          
\hline
 UL            &  22  & 0.52 & 0.55 & 0.22 \\
 He-WD         &  99  & 0.32 & 0.23 & 0.26 \\
 Massive WD    &  29  & 0.21 & 0.20 & 0.15 \\
 NS-NS         &  9   & 0.23 & 0.09 & 0.24 \\
 MS            &  5   & 0.06 & 0.04 & 0.04 \\
 \hline \label{tab:z-stats}               
 \end{tabular}                                                                                 
\end{table}                                                                                    

Table~\ref{tab:z-stats} summarises the statistical distribution of Galactic height for each of the binary groups mentioned above, from which we draw two main interpretations. Firstly, the heavier systems tend to stay closer to the plane, as seen for example from the MS systems with a mean Galactic height of only 0.06\,kpc, whereas the lightest UL systems tend to be found at a higher Galactic height with a mean of 0.52\,kpc. Secondly there is a larger scatter in the height distribution of the lighter systems, whereas the heaviest MS system are found almost exclusively within the Galactic plane. We note that a potential caveat here is that the ages of the MSPs might also have an influence on the Galactic height scattering. For example the fully-recycled He-WD binaries are generally older and hence might have more time to scatter away from the Galactic plane, whereas the less-recycled binaries with heavier companions tend to be younger. In addition, there is also a longer time interval between the SN explosion and the formation of the MSP for systems with UL and with He-WD companions, because their low-mass progenitors have much longer nuclear evolution timescales. Nonetheless, this does not change the outcome of the overall picture in Figure~\ref{fig:mfgb}, explicitly that the distribution of the total mass of binary systems is inversely-related to the Galactic height distribution.

We are aware that the MSP distribution depicted in Figure~\ref{fig:mfgb} is skewed by another observational bias. That is from a pulsar searching point-of-view, pulsars with shorter spin periods, meaning the more recycled UL and He-WD systems, are more difficult to be discovered at higher DM regions, for example deep in the Galactic plane. This is because short spin period pulsars are more vulnerable to dispersion smearing and interstellar scattering. However, the less recycled massive WD and MS systems have longer spin periods, and we should have a relatively more uniform ability to detect them whether they are in the Galactic plane where DM is high or out of the plane. 

This leads to two further implications. The first is that the smaller Galactic heights of the heavier systems are genuine, since if massive WD or MS systems exist at high Galactic heights we would have been more likely to have discovered them, given that we have detected the in-theory more difficult He-WD at those Galactic heights. The second is that this gives an explanation to the lack of light systems at small Galactic heights in the Galactic plane, resulting in the sparsely populated region in Figure~\ref{fig:mfgb} below 0.2\,kpc and for mass function less than $10^{-3}\,M_{\sun}$. Indeed a large number of the UL systems at high Galactic heights are only discovered thanks to the \emph{Fermi} Large Area Telescope \citep[LAT;][]{FermiLAT}, which has much less ability to detect pulsars in the Galactic plane due to confusion with background emission. 

These results show that the observed MSP distribution is not as isotropic as previously thought \citep[see for example][]{Johnston1991} prior to the latest generation of pulsar surveys with improved backends, which have allowed us to probe a much bigger volume within the Galaxy. Conventional MSP population synthesis using the scale factor method typically takes into consideration only the pulsar luminosities \cite[see for example][]{Levin2013}, and we suggest that including the mass function as an extra parameter could be a potential improvement for future population studies. 

\subsection{Orbital eccentricity} \label{sec:Ecc}
\begin{figure}
  \centering
\includegraphics[width=3.4in]{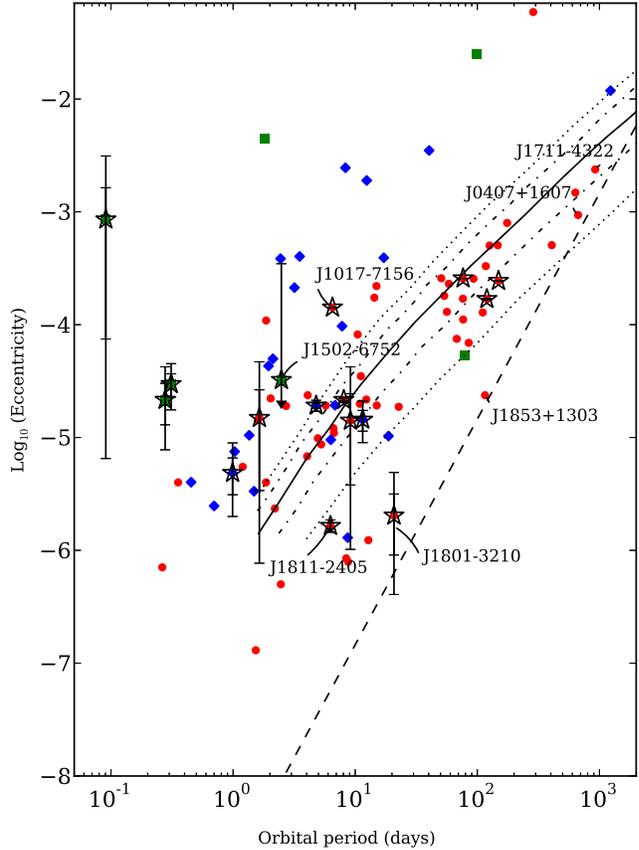}
\caption{Plot of eccentricity vs. orbital period $(P_{\rm{orb}})$. Known pulsars with He-WD companions are plotted as red circles, CO-WD companions in blue diamonds, and ultra-light companions in green squares. The 16 MSPs studied in this work are plotted with star symbols filled with the respective colour according to their companion types, together with the 1- and 2-$\sigma$ uncertainties of the eccentricity measurements. We plot a 2-$\sigma$ upper limit for PSR~J1502$-$6752 where the eccentricity is not constrained. The solid line illustrates the median eccentricity predicted by \protect\citet{Phinney1992}. The dot-dashed line and the dotted line are predicted to contain 68\% and 95\% of the final eccentricities respectively. The dashed line indicates $e \propto {P_{\rm{orb}}}^{2}$.}
\label{fig:PhinSEP}
\end{figure}

We have measured initial eccentricities for the four newly-discovered binary MSPs and improved precision for the eccentricities of the 12 previously published MSPs, except for PSRs~J1502$-$6752 where only upper limits can be achieved. Figure~\ref{fig:PhinSEP} shows a plot of orbital period versus orbital eccentricity and the 16 MSPs in this work are marked together with 1- and 2-$\sigma$ uncertainties of their eccentricities. The dotted lines denote the eccentricity predicted by the convective fluctuation-dissipation theory of \citet{Phinney1992}, applicable to binary systems formed by stable mass transfer from a Roche-lobe filling red giant. It can be seen that our MSPs with He-WD companions (plotted as red stars in Figure~\ref{fig:PhinSEP}) largely agree with the predictions of \citet{Phinney1992}. Within the 2-$\sigma$ eccentricity measurement uncertainties, only PSRs~J1017$-$7156, J1811$-$2405 and J1801$-$3210 lie outside the 95\% confidence-level range (the first one above and the latter two below). However as seen in Figure~\ref{fig:PhinSEP} they have the same scatter as the rest of the MSP population. In addition, these three pulsars have typical He-WD companions and their spin periods indicate highly recycled systems. Therefore, we find little evidence for unusual evolutionary scenarios for these three pulsars.

The low eccentricity of $e=2.1\pm1.1\times10^{-6}$ of PSR~J1801$-$3210 combined with its large orbital period of $P_{\rm{orb}} = 21$\,days makes it a `wide-orbit binary millisecond pulsar' (WBMSP), and an interesting object to be employed for tests of the strong equivalence principle (SEP) as described in \citet{DS1991,Stairs2005,Freire2012}. The basic idea being that in the case of SEP violation, the extreme difference between the gravitational binding energy of the heavy neutron star and its much less compact companion star implies that they would experience different accelerations in the presence of an external gravitational field (Nordvedt effect). This translates to an observable effect, most prominent in systems with small eccentricity and wide orbits, that the eccentricity would oscillate between the minimum and maximum value. The dashed line overplotted on Figure~\ref{fig:PhinSEP} indicates $e \propto {P_{\rm{orb}}}^{2}$, a figure-of-merit for a SEP test. With a ${P_{\rm{orb}}}^{2}/e$ ratio of $2.1\times10^{8}\rm\,day^{2}$, PSR~J1801$-$3210 thus provides the best test for SEP together with PSRs~J1835$+$1303 and J0407$+$1607 as detailed in \citet{Gonzalez2011}. Note that although PSR~J1711$-$4322 appears to lie close to the figure-of-merit in Figure~\ref{fig:PhinSEP}, it is in fact not usable for this SEP test \citep{KK2012}. 

\subsection{Change in projected semi-major axis, $\dot{x}$} \label{sec:xdot}
For PSR~J1017$-$7156 we determine a change in projected semi-major axis $(\dot{x})$ of $9.1\pm1.7\times10^{-15}$. The projected semi-major axis, $x$, is related to the semi-major axis, $a_{\rm{p}}$, and the inclination, $i$, by Equation~(\ref{eq:x}). Hence a measurement of $\dot{x}$ could be due either to a physical change of the intrinsic orbit size as measured by $a_{\rm{p}}$, or to a change in $i$, or both. 

In the case of an actual change in $a_{\rm{p}}$ due to gravitational wave emission, we would expect this to also be reflected in a detection of $\dot{P}_{\rm{orb}}$ \citep{Peters1964}. From this we can predict the corresponding observable change in $a_{\rm{p}}\sin{i}/c$ to be of the order of $10^{-21}$ for PSR~J1017$-$7156, which is many orders of magnitude too small to be observed. So we conclude that the observed $\dot{x}$ is most likely due to an apparent change in the orbital inclination as a result of proper motion affecting the viewing geometry. This effect has been first proposed by \citet{Arzoumanian1996} and \citet{Kopeikin1996} using,
\begin{equation}
\dot{x} = 1.54\times10^{-16} x \left(\frac{\mu}{\rm mas\,yr^{-1}}\right) \cot i~ \sin ( \Theta - \Omega )\,.
\label{eq:xdot}
\end{equation}
In this equation proper motion has a total magnitude of $\mu$ and a position angle of $\Theta$, whereas $\Omega$ is the position angle of the line of nodes.  

To assess if any physical constraints of the orientation of the line of nodes in relation to the direction of the proper motion (i.e. $\Theta\,-\,\Omega$) can be subsequently drawn, one must compare the uncertainty of the measured $\dot{x}$ with the product of $\mu$ and $x$. For PSR~J1017$-$7156 we have $\mu{x} = 7.6\times10^{-15}$, which is indeed in the same order of magnitude as compared to our $\dot{x}$ measurement, and can already provide constraints to the possible ranges of $\Omega$. Future improved timing precision and additional information, such as constraints on or detection of a Shapiro delay, will allow us to extract more information on the binary systems, including mass measurements. None of the other MSPs in this paper have a detectable $\dot{x}$ yet and are unlikely to be measurable in the near future. With the possible exceptions of PSRs~J1125$-$5825 and J1708$-$3506, which both have $\mu$$x$ of the order of $10^{-14}$, we can quote a marginal $\dot{x}$ limit of $1.6\pm2.0\times10^{-14}$ and $-9\pm6\times10^{-14}$ respectively. Hence they might achieve reliable $\dot{x}$ measurements with additional timing data. 

\subsection{Orbital period variation, $\dot{P}_{\rm{orb}}$} \label{sec:Pbdot}
We measure an orbital period variation $(\dot{P}_{\rm{orb}})$ in PSR~J1731$-$1847. However rather than due to gravitational-wave damping, the $\dot{P}_{\rm{orb}}$ observed in this case is more likely due to the eclipsing nature of PSR~J1731$-$1847, a BW system, inducing orbital interaction. We refer to \citet{Lazaridis2011} for a detailed discussion of such orbital period variations caused by changes in the gravitational quadrupole moment of a tidally interacting BW system. For the case of PSR~J1731$-$1847, a straight-forward fit of $\dot{P}_{\rm{orb}}$ is not adequate, since the orbital period exhibits quadratic changes over the last three years. We have achieved the best fit using the BTX model (Nice, D., unpublished) implemented in \textsc{tempo2}, taking into account the orbital frequency changes up to the second order term (i.e. ${n_{\rm{b}}}$, $\dot{n}_{\rm{b}}$, $\ddot{n}_{\rm{b}}$). The phase $(\phi)$ of the orbit is thus a function of the binomial expansion of the $n_{\rm{b}}^{(k)}$ terms, where $k$ denotes the ${k^{\rm{th}}}$ derivative with respect to time. At any particular time, $t$, the phase $\phi$ can be represented by,

\begin{equation}
\phi(t) = \sum_{k=1}^{K} \left(\frac{n_{\rm{b}}^{(k)}}{k+1!}\right) \left(\frac{t-T_{0}}{s}\right)^{k+1} / n_{\rm{b}}\,.
\end{equation}

To get a better visualisation of the change of the orbit over time, we express this phase shift as the shift of the epoch of periastron ($T_{0}$). One can consider that a positive phase shift corresponds to an earlier arrival of the observed periastron, $T_{\rm{0,obs}}$, as compared to the predicted arrival of the periastron, $T_{\rm{0,pre}}$. The result is a negative $\Delta T_{0}$, which also symbolises a decrease in $\dot{P}_{\rm{orb}}$,

\begin{equation}
\Delta T_{0} = T_{\rm{0,pre}} - T_{\rm{0,obs}} = \Delta \phi \times P_{\rm{orb}}\,.  
\end{equation}

\begin{figure}
\includegraphics[width=3.4in]{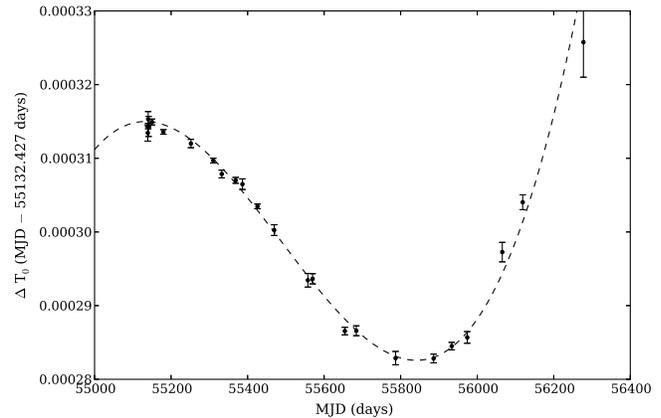}
\caption{A plot of $\Delta T_{0}$ as a function of time for PSR~J1731$-$1847. The dashed line shows the best-fit curve from the timing solution generated with the BTX model in \textsc{tempo2}, employing up to the second orbital frequency derivative terms.} \label{fig:eclipseT}
\end{figure}

Figure \ref{fig:eclipseT} shows this $\Delta T_{0}$ as derived from the $n_{\rm{b}}^{(k)}$ terms of the BTX model fit in \textsc{tempo2}. It can be seen that the orbit of PSR~J1731$-$1847 shrinks until approximately MJD 55800 but gets wider after. We identified manually a value of $T_{0}$ across every few TOAs, while holding fixed all other parameters (shown by black points in Figure~\ref{fig:eclipseT}). The BTX model results in a close agreement. We remark, however, that this model has no predictive power for the orbital period variations outside of the current TOA timeline.

\subsection{Variation in the longitude of periastron, $\dot{\omega}$} \label{sec:PK}
We measured a marginally significant variation in the longitude of periastron $(\dot{\omega})$ for PSR~J1017$-$7156 with a value of $0.022\pm0.009\,\degr\rm\,yr^{-1}$. If we assume a typical pulsar mass of 1.4$\,M_{\sun}$ and an orbital inclination of 60$\,\degr$, using Equation~(2) of \citet{Weisberg1981} we obtain a predicted $\dot{\omega}$ in general relativity of 0.012$\,\degr\rm\,yr^{-1}$, which agrees with our measured value within 1.1-$\sigma$. In general $\dot{\omega}$ is a useful Post-Keplerian (PK) parameter as it can be used to calculate the total mass of the binary system, from which a measurement of the pulsar mass may be extracted. The variation in $\dot{\omega}$ is the easiest to measure for orbits with significant eccentricities. In the case of PSR~J1017$-$7156 with $e = 1.4\times10^{-4}$ and an already good timing residual RMS of 1.3 $\rm\mu$s, we expect its $\dot{\omega}$ measurement to be much improved with another 5 years of timing data.

\subsection{Gamma-ray pulsation searches} \label{sec:gamma}
Among the pulsars in our sample, PSRs~J1125$-$5825 and J1446$-$4701 have been observed to emit $>0.1$\,GeV pulsations by \citet{HTRU4}, through the analysis of data taken by the \emph{Fermi} Large Area Telescope \citep[LAT;][]{FermiLAT}, with post-trial significances just under 5-$\sigma$. High confidence detections of these two MSPs in gamma rays were later presented in \citet{Fermi2PC}.

\begin{table*}
\begin{center}
\caption{Gamma-Ray Emission Properties of PSRs~J1125$-$5825, J1446$-$4701, J1543$-$5149, and J1811$-$2405. The weighted $H$-test parameters were calculated by selecting photons found within 5$^\circ$ of the pulsars, with energies larger than 0.1\,GeV and weights larger than 0.01. See Figure~\ref{fig:lightcurves} for the corresponding gamma-ray light curves under the same selection cuts. Details on the measurement of the spectral parameters can be found in Section~\ref{sec:gamma}.}
\begin{tabular}{lcccc}
\hline
Parameter & J1125$-$5825 & J1446$-$4701 & J1543$-$5149 & J1811$-$2405\\
\hline
Weighted $H$-test & 
100.7 & 
165.4 & 
65.1 & 
37.9 \\

Spectral index, $\Gamma$ & 
$1.6 \pm 0.5$ & 
$1.3 \pm 0.4$ & 
$2.3 \pm 0.3$ & 
$1.6 \pm 0.4$ \\

Cutoff energy, $E_c$ (GeV) & 
$8 \pm 7$ & 
$4 \pm 2$ & 
$6 \pm 3$ & 
$3 \pm 2$ \\

Photon flux above 100\,MeV, $F_{100}$ $(10^{-8}\rm\,cm^{-2}\rm\,s^{-1})$ & 
$0.8 \pm 0.7$ & 
$0.6 \pm 0.2$ & 
$5.4 \pm 0.4$ & 
$2 \pm 2$ \\

Energy flux above 100\,MeV, $G_{100}$ $(10^{—11}\rm\,erg\rm\,cm^{-2}\rm\,s^{-1})$ & 
$0.9 \pm 0.3$ & 
$0.7 \pm 0.1$ & 
$2.4 \pm 0.2$ & 
$1.4 \pm 0.8$ \\

Luminosity, $L_\gamma = 4 \pi G_{100} d^2$ $(10^{33}\rm\,erg\rm,s^{-1})$ & 
$7.1 \pm 2.4$ & 
$1.9 \pm 0.3$ & 
$17 \pm 1$ & 
$5.5 \pm 3.1$ \\

Efficiency, $\eta = L_\gamma / \dot E$ & 
$0.09 \pm 0.03$ & 
$0.05 \pm 0.01$ & 
$0.23 \pm 0.02$ & 
$0.2  \pm 0.1$ \\

\hline \label{tab:Fermispec}
\end{tabular}
\end{center}
\end{table*}

\begin{figure*}
\begin{center}
\includegraphics[scale=0.67]{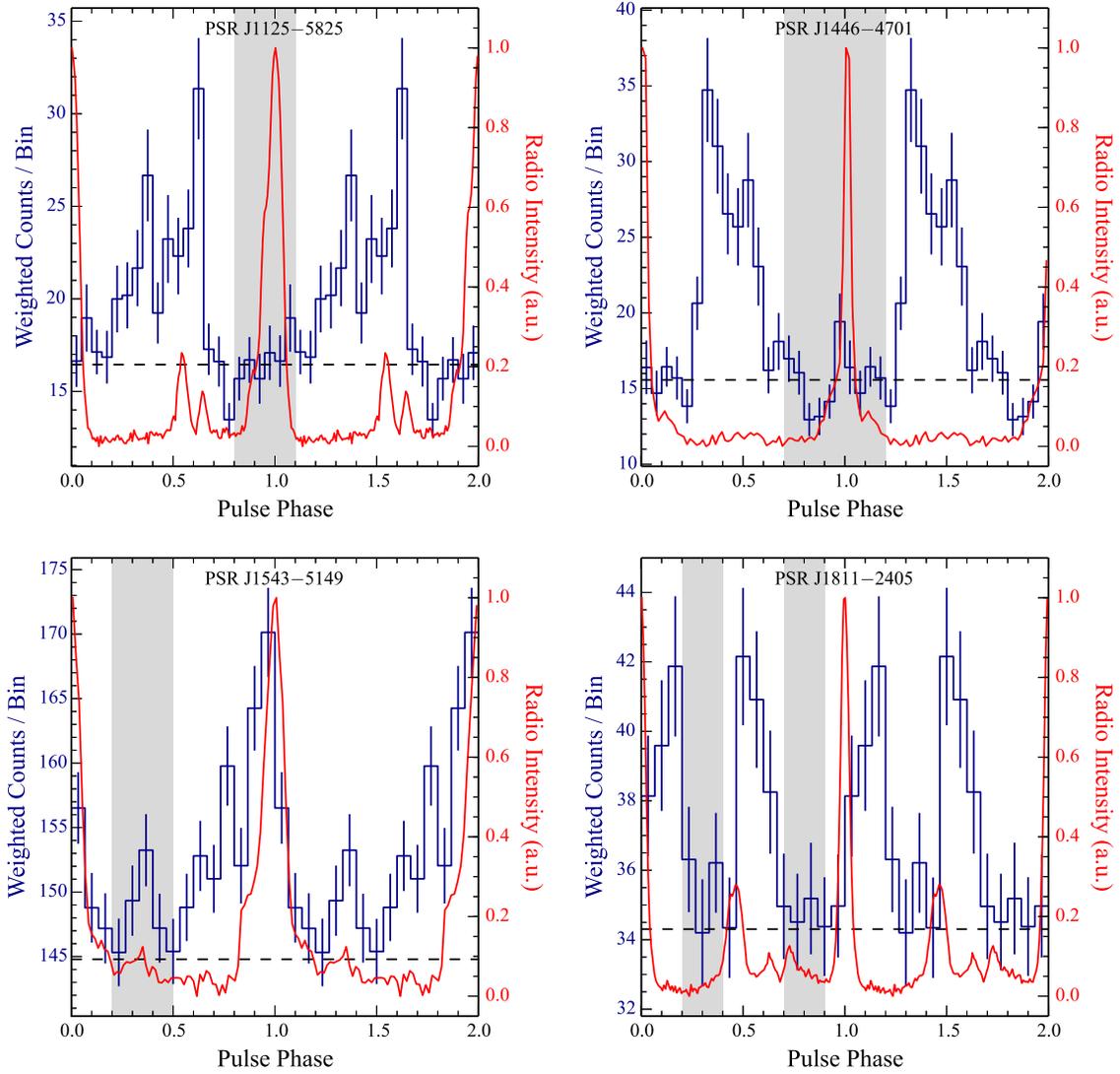}
\caption{Radio and gamma-ray light curves for the four MSPs in our sample with \textit{Fermi} LAT detections. Two pulsar cycles are shown for clarity. The radio profiles are based on 1.4 GHz observations conducted at Parkes, while the gamma-ray profiles were obtained by selecting \textit{Fermi} LAT photons with reconstructed directions found within $5^\circ$ of the MSPs, and with energies larger than 0.1\,GeV. The photons were weighted by the probability that they originate from the pulsars as described in e.g. \protect \citet{Kerr2011}. Photons with weights smaller than 0.01 were rejected. Horizontal dashed lines show the estimated background levels, obtained by following the method described in \citet{Guillemot2012}. The grey shaded regions indicate the OFF-pulse intervals used for the spectral analyses presented in Section~\ref{sec:gamma}, the ON-pulse regions being defined as the complementary intervals. \label{fig:lightcurves}}
\end{center}
\end{figure*}

To determine whether other MSPs in our sample also emit gamma-ray pulsations, we analysed LAT photons recorded between \fermimin{} and \fermimax{}, with energies from 0.1 to 100\,GeV, and belonging to the `Source' class of the reprocessed P7REP data, a version of Pass7 data\footnote{See \protect \citet{Bregeon2013} and http://fermi.gsfc.nasa.gov/ssc/data/analysis/documentation/Pass7REP\_usage.html for more information.} reprocessed with improved calibrations. Events with zenith angles larger than $100\degr{}$ were excluded, to reject atmospheric gamma rays from the Earth's limb. In addition, events recorded when the instrument was not operating in nominal science operations mode, when the limb of the Earth infringed upon the regions of interest (see below for the definition of these regions), or when the data were not flagged as good were excluded. These cuts were made using the \emph{Fermi} Science Tools\footnote{http://fermi.gsfc.nasa.gov/ssc/data/analysis/scitools/overview.html} (STs) v9r32p5, and the selected photons were assigned pulse phases using the ephemerides listed in Tables~\ref{tab:ell1} and \ref{tab:dd} and the \emph{Fermi} plug-in for \textsc{tempo2} \citep{Ray2011,Hobbs2006}. 

Weighting each event by the probability that it originates from a pulsar has been shown to make pulsation searches more sensitive \citep[e.g.,][]{Kerr2011}. We calculated these weights by performing binned maximum likelihood analyses for each pulsar, using the \textit{pyLikelihood} python module distributed with the STs. For each MSP we selected photons found in a region of radius 15$^\circ$ centred on the pulsar, and built a spectral model for this region by including sources within $20^\circ$, from a preliminary list based on four years of LAT data. The Galactic diffuse emission was modelled using the \textit{gll\_iem\_v05.fit} map, and the isotropic diffuse emission and residual instrumental background were modelled using the \textit{iso\_source\_v05.txt} template. We used the P7REP\_SOURCE\_V15 instrument response functions, and followed the analysis prescriptions described in \citet{Fermi2PC}. However, in a first iteration of the analysis the MSPs were modelled with simple power laws of the form $N_0 \left(E / \mathrm{GeV}\right)^{-\Gamma}$, where $N_0$ is a normalisation factor, $E$ denotes the photon energy and $\Gamma$ the photon index. A test statistic \citep[see][for a definition]{Fermi2FGL} larger than 40 was found for PSRs~J1125$-$5825, J1446$-$4701, J1543$-$5149, and J1811$-$2405, indicating the presence of significant gamma-ray emission. No evidence for gamma-ray emission from any of the other pulsars was found. For these pulsars, we have conducted an unweighted search for pulsations, testing a range of angular and energy cuts to the LAT data to optimise the $H$-test statistic \citep{deJager2010}. We did not find evidence for gamma-ray pulsations with significance greater than 3-$\sigma$ in any of the data selection cuts used, for these pulsars without significant continuous emission.

For the four MSPs with gamma-ray detections, we computed the weights using the ST \texttt{gtsrcprob} and the best-fit spectral models as obtained from the preliminary likelihood analyses. For PSRs~J1125$-$5825, J1446$-$4701, and J1543$-$5149, we found spectrally-weighted $H$-test significances \citep{Kerr2011} above 5-$\sigma$, while for J1811$-$2405 we obtained a 4.4-$\sigma$ detection, suggesting that J1543$-$5149 and J1811$-$2405 are indeed gamma-ray pulsars. In order to improve the quality of the spectral results and thereby increase the weighted pulsation significances we inspected the preliminary light curves for the four MSPs visually to determine ON-pulse regions, that we refit with \texttt{gtlike}, this time modelling the MSPs with exponentially cutoff power laws of the form $N_0 \left(E / \mathrm{GeV}\right)^{-\Gamma} \exp \left(-E/E_c \right)$, where $E_c$ is the cutoff energy. The best-fit spectral parameters obtained from this second iteration are listed in Table~\ref{tab:Fermispec}, and the spectrally-weighted light curves are shown in Figure~\ref{fig:lightcurves} along with the ON-pulse intervals chosen for this analysis. For all four pulsars, the $H$-test parameters using a minimum weight cut of 0.01 all indicate $>5$-$\sigma$ detections, even after accounting for the trial factor due to the two analysis steps. 

The spectral parameters listed in Table~\ref{tab:Fermispec} for PSRs~J1125$-$5825 and J1446$-$4701 are consistent with those reported in \citet{Fermi2PC} to within uncertainties. The parameters for PSRs~J1543$-$5149 and J1811$-$2405 are only weakly constrained at present, but are reminiscent of those of known gamma-ray MSPs \citep{Fermi2PC}. Also listed in Table~\ref{tab:Fermispec} are the gamma-ray luminosities $L_\gamma$ deduced from the energy flux measurements, and the efficiencies of conversion of spin-down power into gamma-ray emission, $\eta = L_\gamma / \dot E$, calculated using the Shklovskii-corrected $\dot E$ values and the DM distances given in Tables~\ref{tab:ell1} and \ref{tab:dd}. The uncertainties reported in Table~\ref{tab:Fermispec} are statistical. Studies of systematic uncertainties in the effective area suggest a 10\% uncertainty at 100\,MeV, decreasing linearly in Log(E) to 5\% in the range between 316\,MeV and 10\,GeV and increasing linearly in Log(E) up to 10\% at 1\,TeV\footnote{see http://fermi.gsfc.nasa.gov/ssc/data/analysis/LAT\_caveats.html}. 

The two newly-identified gamma-ray pulsars, PSRs~J1543$-$5149 and J1811$-$2405, bring the total number of MSPs with detected gamma-ray pulsations to 53 objects. It is unlikely that the \emph{Fermi} LAT will detect many of the remaining MSPs presented in this paper. Assuming an average gamma-ray efficiency for MSPs of 0.245 following the prescription of \citet{FermiB1821}, we derive expected energy fluxes for these pulsars much smaller than the lowest value reported in \citet{Fermi2PC} for an MSP, because of the generally large distance values; with the notable exception of PSR~J1731$-$1438. The latter MSP may be inefficient at converting its spin-down power into gamma-ray emission, or its gamma-ray beams may not cross the Earth's line-of-sight. The high $\dot E$ but distant MSPs in this sample could contribute to the diffuse emission seen by the \emph{Fermi} LAT around the Galactic plane.

\section{Conclusion} \label{sec:conclusion}
The High Time Resolution Universe survey for pulsars and fast transients has discovered 27 MSPs to date, of which four are announced in this work. All four MSPs have phase-coherent timing solution with RMS already of the order of tens of $\mu$s. We have presented their pulse profiles and polarimetric properties at different frequencies. PSRs~J1528$-$3828 and J1056$-$7117 are likely to be formed from wide-orbit LMXBs, leading to the formation of classic MSPs with He-WD companions. PSR~J1755$-$3716 is likely to have evolved from an IMXB and possesses a CO-WD companion. PSR~J1525$-$5545 is likely to have a massive CO-WD companion, or an ONeMg-WD companion if the orbital inclination angle is low. 

In addition, we present updated timing solutions for 12 previously published HTRU MSPs, as compared with results in their respective discovery papers (\cite{HTRU2}, \cite{HTRU4}, \cite{Bailes2011}), thanks to the now longer timing baseline of over three years in all cases, except one with 2.7\,years of timing data.

We measure 5 new proper motions with significance greater than 3-$\sigma$, from PSRs~1017$-$7156, J1125$-$5825, J1446$-$4701, J1708$-$3506 and J1719$-$1438. Their derived transverse velocities are all consistent with previous MSP velocity distribution studies. In turn, with the proper motion measurements, we are able to constrain the period derivative contribution from the Shklovskii effect. In addition, we take into account the acceleration due to the Galactic potentials and correct for the intrinsic period derivatives for the 12 MSPs in this work. PSR~J1017$-$7156 has one of the smallest inferred intrinsic period derivatives at 1.2$\times10^{-21}$, hence also one of the lowest derived surface magnetic field strength within the known MSP population at a value of $5.4\times10^{7}$\,G. 

We further discuss the case of PSR~J1801$-$3210 for which no significant period derivative can be measured, even with more than 4 years of timing data. The best-fit solution in \textsc{tempo2} shows a $\dot{P}_{\rm{obs}}$ of $-4\pm4\times10^{-23}$, an extremely small number comparing to that of a typical MSP. The both positive $\dot{P}_{\rm{shk}}$ and $\dot{P}_{\rm{gal}}$ of the order of $10^{-20}$ and $10^{-21}$, respectively, act to further decrease the already negative period derivative. It seems unlikely that the DM-derived distance is significantly wrong and hence reversing the direction of Galactic potential. Alternatively we consider the presence of a third body near PSR~J1801$-$3210 which might be accelerating the pulsar towards Earth. Giant molecular clouds seem to be a plausible scenario, whereas an exoplanet orbiting in a large hierarchical orbit seems unlikely due to the small probability of surviving the SN, as well as the fact that we do not measure any significant second derivatives of spin frequency. Based on radio emission theory, we derive a theoretical lower limit of period derivative of $7.9\times10^{-24}$ and a corresponding surface magnetic field strength at the equator of $7.8\times10^{6}$\,G for PSR~J1801$-$3210, in order for it to stay above the pulsar death line. We also highlight the potential of PSR~J1801$-$3210 to be employed in the SEP test due to its wide and circular orbit. 

We have undertaken a comparison study between MSPs in our sample and the complete known pulsar population. We point to a strong dependence on inclination for eclipses to be observed in VLMBPs, as indicated by an apparent bimodal distribution of eclipsing and non-eclipsing systems separated by a companion mass of about $0.027\,M_{\sun}$. We also suggest that the distribution of the total mass of binary systems is inversely-related to the Galactic height distribution. In other words, MSPs with the heaviest companions have larger tendencies to stay close to the Galactic plane, whereas lighter systems with smaller mass functions show larger mean value and larger scatter in the Galactic height distribution. 

A change in the projected semi-major axis ($\dot{x}$) is observed in PSR~J1017$-$7156 at $9.1\pm1.7\times10^{-15}$. Rather than due to gravitational wave emission, this $\dot{x}$ is likely due to an apparent change in the orbital inclination as a result of proper motion affecting the viewing geometry. We also report an $\dot{\omega}$ of 0.022(9)$\,\degr\rm\,yr^{-1}$, and we highlight the potential of measuring more relativistic orbital parameters with PSR~J1017$-$7156. Together with its small period derivative and the corresponding low derived magnetic field as mentioned above, this makes PSR~J1017$-$7156 a very interesting pulsar to be closely followed-up with further timing campaign, and indeed it is already being monitored by the Parkes Pulsar Timing Array Project \citep{Manchester2013}. Although we stress the importance of a careful dispersion measure variation treatment as discussed in Section~\ref{sec:DM} and a proper polarisation calibration to correctly assess the uncertainties on the TOAs for the high-precision timing required for PSR~J1017$-$7156. Furthermore, orbital period variations are observed in the BW system PSR~J1731$-$1847. We present the timing solution with the BTX timing model which demonstrate the quadratic changes in orbital period over the last three years.  

We detected highly significant gamma-ray pulsations from PSRs~J1125$-$5825 and J1446$-$4701, confirming the results of \citet{HTRU4} and \citet{Fermi2PC}. PSRs~J1543$-$5149 and J1811$-$2405 were identified for the first time as gamma-ray pulsars: after folding the \textit{Fermi} LAT photons with radio timing ephemerides, we obtained $>$ 5-$\sigma$ detections of these two MSPs, bringing the total number of MSPs with detected gamma-ray pulsations to 53 objects. 

\section*{acknowledgements}
The Parkes Observatory is part of the Australia Telescope, which is funded by the Commonwealth of Australia for operation as a National Facility managed by CSIRO. A large amount of the timing data for PSR~J1017$-$7156 has been taken as part of the Parkes Pulsar Timing Array Project P456.

The \textit{Fermi} LAT Collaboration acknowledges generous ongoing support from a number of agencies and institutes that have supported both the development and the operation of the LAT as well as scientific data analysis. These include the National Aeronautics and Space Administration and the Department of Energy in the United States, the Commissariat \`a l'Energie Atomique and the Centre National de la Recherche Scientifique / Institut National de Physique Nucl\'eaire et de Physique des Particules in France, the Agenzia Spaziale Italiana and the Istituto Nazionale di Fisica Nucleare in Italy, the Ministry of Education, Culture, Sports, Science and Technology (MEXT), High Energy Accelerator Research Organization (KEK) and Japan Aerospace Exploration Agency (JAXA) in Japan, and the K.~A.~Wallenberg Foundation, the Swedish Research Council and the Swedish National Space Board in Sweden. Additional support for science analysis during the operations phase is gratefully acknowledged from the Istituto Nazionale di Astrofisica in Italy and the Centre National d'\'Etudes Spatiales in France.

The authors would like to thank Patrick Lazarus for providing an automated RFI cleaning routine, Dominic Schnitzeler for useful discussions on pulsar distances, Joris Verbiest for his expertise on the use of \textsc{tempo2}, Charlotte Sobey and Gregory Desvignes for advices on data calibration, Kejia Lee for his help in plotting Figure~\ref{fig:PhinSEP}, Chris Flynn for sharing his knowledge on Galactic stellar density, Megan DeCesar from the \textit{Fermi} collaboration for reviewing the paper and providing many constructive suggestions, and Lijing Shao for carefully reading the manuscripts. CN was supported for this research through a stipend from the International Max Planck Research School (IMPRS) for Astronomy and Astrophysics at the Universities of Bonn and Cologne.

\bibliographystyle{mn2e}

\begin{thebibliography}{}
\bibitem[\protect\citeauthoryear{{Abdo} et~al.,}{{Abdo} et~al.}{2009}]{Abdo2009}
{Abdo} A.~A. {\rm et~al.}, 2009, Science, 325, 848

\bibitem[\protect\citeauthoryear{{Abdo} et~al.,}{{Abdo} et~al.}{2013}]{Fermi2PC}
{Abdo} A.~A. {\rm et~al.}, 2013, \apjs, 208, 17

\bibitem[\protect\citeauthoryear{{Alpar}, {Cheng}, {Ruderman} \& {Shaham}}{{Alpar} et~al.}{1982}]{Alpar1982}
{Alpar} M.~A.,  {Cheng} A.~F.,  {Ruderman} M.~A.,    {Shaham} J.,  1982, \nat, 300, 728

\bibitem[\protect\citeauthoryear{{Antoniadis} et~al.,}{{Antoniadis} et~al.}{2013}]{Antoniadis2013}
{Antoniadis} J. {\rm et~al.}, 2013, Science, 340, 448

\bibitem[\protect\citeauthoryear{{Arzoumanian}, {Joshi}, {Rasio} \& {Thorsett}}{{Arzoumanian} et~al.}{1996}]{Arzoumanian1996}
{Arzoumanian} Z.,  {Joshi} K.,  {Rasio} F.~A.,    {Thorsett} S.~E.,  1996, in {Johnston} S.,  {Walker} M.~A.,   {Bailes} M.,  eds, IAU Colloq. 160:
  Pulsars: Problems and Progress Vol.~105 of Astronomical Society of the Pacific Conference Series, {Orbital Parameters of the PSR~B1620-26 Triple
  System}. 
pp 525--530

\bibitem[\protect\citeauthoryear{{Atwood}, {Abdo}, {Ackermann}, {Althouse}, {Anderson}, {Axelsson}, {Baldini}, {Ballet}, {Band}, {Barbiellini} \& et
  al.}{{Atwood} et~al.}{2009}]{FermiLAT}
{Atwood} W.~B. {\rm et~al.}, 2009, \apj, 697, 1071

\bibitem[\protect\citeauthoryear{{Backer}, {Foster} \& {Sallmen}}{{Backer} et~al.}{1993}]{Backer1993}
{Backer} D.~C.,  {Foster} R.~S.,    {Sallmen} S.,  1993, \nat, 365, 817

\bibitem[\protect\citeauthoryear{{Bailes}, {Bates}, {Bhalerao}, {Bhat}, {Burgay}, {Burke-Spolaor}, {D'Amico}, {Johnston}, {Keith}, {Kramer},
  {Kulkarni}, {Levin}, {Lyne}, {Milia}, {Possenti}, {Spitler}, {Stappers} \& {van Straten}}{{Bailes} et~al.}{2011}]{Bailes2011}
{Bailes} M. {\rm et~al.}, 2011, Science, 333,
  1717

\bibitem[\protect\citeauthoryear{{Barr}, {Champion}, {Kramer}, {Eatough}, {Freire}, {Karuppusamy}, {Lee}, {Verbiest}, {Bassa}, {Lyne}, {Stappers},
  {Lorimer} \& {Klein}}{{Barr} et~al.}{2013}]{Barr2013}
{Barr} E.~D. {\rm et~al.},  2013, \mnras

\bibitem[\protect\citeauthoryear{{Bhattacharya}}{{Bhattacharya}}{2002}]{Bhattacharya2002}
{Bhattacharya} D.,  2002, Journal of Astrophysics and Astronomy, 23, 67

\bibitem[\protect\citeauthoryear{{Bates}, {Bailes}, {Bhat}, {Burgay}, {Burke-Spolaor}, {D'Amico}, {Jameson}, {Johnston}, {Keith}, {Kramer},
  {Levin}, {Lyne}, {Milia}, {Possenti}, {Stappers} \& {van Straten}}{{Bates} et~al.}{2011}]{HTRU2}
{Bates} S.~D. {\rm et~al.}, 2011, \mnras, 416, 2455

\bibitem[\protect\citeauthoryear{{Beskin}, {Gurevich} \& {Istomin}}{{Beskin} et~al.}{1988}]{Beskin1988}
{Beskin} V.~S.,  {Gurevich} A.~V.,    {Istomin} I.~N.,  1988, \apss, 146, 205

\bibitem[\protect\citeauthoryear{{Bisnovatyi-Kogan} \& {Komberg}}{{Bisnovatyi-Kogan} \& {Komberg}}{1974}]{BK1974}
{Bisnovatyi-Kogan} G.~S.,  {Komberg} B.~V.,  1974, \sovast, 18, 217

\bibitem[\protect\citeauthoryear{{Bovy} \& {Rix}}{{Bovy} \& {Rix}}{2013}]{Bovy2013}
{Bovy} J.,  {Rix} H.-W.,  2013, ArXiv e-prints

\bibitem[\protect\citeauthoryear{{Bregeon}, {Charles} \& {M.~Wood for the Fermi-LAT collaboration}}{{Bregeon} et~al.}{2013}]{Bregeon2013}
{Bregeon} J.,  {Charles} E.,    {M.~Wood for the Fermi-LAT collaboration} 2013, {Proceedings of the 4th \textit{Fermi} symposium, eConf C121028
  (arXiv:1304.5456)}

\bibitem[\protect\citeauthoryear{{Champion} et~al.,}{{Champion} et~al.}{2008}]{Champion2008}
{Champion} D.~J. {\rm et~al.}, 2008, Science, 320, 1309

\bibitem[\protect\citeauthoryear{{Champion} et~al.,}{{Champion} et~al.}{2010}]{Champion2010}
{Champion} D.~J. {\rm et~al.}, 2010, \apjl, 720, L201

\bibitem[\protect\citeauthoryear{{Chen}, {Chen}, {Tauris} \& {Han}}{{Chen} et~al.}{2013}]{Chen2013}
{Chen} H.-L.,  {Chen} X.,  {Tauris} T.~M.,    {Han} Z.,  2013, \apj, 775, 27

\bibitem[\protect\citeauthoryear{{Chen} \& {Ruderman}}{{Chen} \& {Ruderman}}{1993}]{Chen1993}
{Chen} K.,  {Ruderman} M.,  1993, \apj, 402, 264

\bibitem[\protect\citeauthoryear{{Cordes} \& {Lazio}}{{Cordes} \& {Lazio}}{2002}]{NE2001model}
{Cordes} J.~M.,  {Lazio} T.~J.~W.,  2002, ArXiv Astrophysics e-prints

\bibitem[\protect\citeauthoryear{{Cumming}, {Zweibel} \& {Bildsten}}{{Cumming} et~al.}{2001}]{C2001}
{Cumming} A.,  {Zweibel} E.,    {Bildsten} L.,  2001, \apj, 557, 958

\bibitem[\protect\citeauthoryear{{Damour} \& {Deruelle}}{{Damour} \& {Deruelle}}{1986}]{DD1986}
{Damour} T.,  {Deruelle} N.,  1986, Ann.~Inst.~Henri Poincar{\'e} Phys.~Th{\'e}or., Vol.~44, No.~3, p.~263 - 292, 44, 263

\bibitem[\protect\citeauthoryear{{Damour} \& {Sch\"afer}}{{Damour} \& {Sch\"afer}}{1991}]{DS1991}
{Damour} T.,  {Sch\"afer} G.,  1991, Physical Review Letters, 66, 2549

\bibitem[\protect\citeauthoryear{{Damour} \& {Taylor}}{{Damour} \& {Taylor}}{1991}]{DT1991}
{Damour} T.,  {Taylor} J.~H.,  1991, \apj, 366, 501

\bibitem[\protect\citeauthoryear{{de Jager} \& {B{\"u}sching}}{{de Jager} \& {B{\"u}sching}}{2010}]{deJager2010}
{de Jager} O.~C.,  {B{\"u}sching} I.,  2010, \aap, 517, L9

\bibitem[\protect\citeauthoryear{{Demorest}, {Pennucci}, {Ransom}, {Roberts} \& {Hessels}}{{Demorest} et~al.}{2010}]{Demorest2010}
{Demorest} P.~B.,  {Pennucci} T.,  {Ransom} S.~M.,  {Roberts} M.~S.~E., {Hessels} J.~W.~T.,  2010, \nat, 467, 1081

\bibitem[\protect\citeauthoryear{{Espinoza} et~al.,}{{Espinoza} et~al.}{2013}]{Espinoza2013}
{Espinoza} C.~M. {\rm et~al.}, 2013, \mnras, 430, 571

\bibitem[\protect\citeauthoryear{{Folkner}, {Williams} \& {Boggs}}{{Folkner} et~al.}{2009}]{DE421}
{Folkner} W.~M.,  {Williams} J.~G.,    {Boggs} D.~H.,  2009, Interplanetary Network Progress Report, 178, C1

\bibitem[\protect\citeauthoryear{{Freire}}{{Freire}}{2005}]{Freire2005}
{Freire} P.~C.~C.,  2005, in {Rasio} F.~A.,  {Stairs} I.~H.,  eds, Binary Radio Pulsars Vol.~328 of Astronomical Society of the Pacific Conference Series,
  {Eclipsing Binary Pulsars}. p.~405

\bibitem[\protect\citeauthoryear{{Freire}, {Wex}, {Esposito-Far{\`e}se}, {Verbiest}, {Bailes}, {Jacoby}, {Kramer}, {Stairs}, {Antoniadis} \&
  {Janssen}}{{Freire} et~al.}{2012}]{Freire2012}
{Freire} P.~C.~C. {\rm et~al.},  2012, \mnras, 423, 3328

\bibitem[\protect\citeauthoryear{{Gonzalez}, {Stairs}, {Ferdman}, {Freire}, {Nice}, {Demorest}, {Ransom}, {Kramer}, {Camilo}, {Hobbs}, {Manchester} \&
  {Lyne}}{{Gonzalez} et~al.}{2011}]{Gonzalez2011}
{Gonzalez} M.~E. {\rm et~al.},  2011, \apj, 743, 102

\bibitem[\protect\citeauthoryear{{Granet}, {Zhang}, {Forsyth}, {Graves}, {Doherty}, {Greene}, {James}, {Sykes}, {Bird}, {Sinclair}, {Moorey} \&
  {Manchester}}{{Granet} et~al.}{2005}]{1050CM}
{Granet} C. {\rm et~al.},  2005, IEEE Antennas Propagation Magazine, 47, 13

\bibitem[\protect\citeauthoryear{{Guillemot} et~al.,}{{Guillemot} et~al.}{2012}]{Guillemot2012}
{Guillemot} L. {\rm et~al.}, 2012, \apj, 744, 33

\bibitem[\protect\citeauthoryear{{Hills}}{{Hills}}{1983}]{Hills1983}
{Hills} J.~G.,  1983, \apj, 267, 322

\bibitem[\protect\citeauthoryear{{Hobbs} et~al.,}{{Hobbs} et~al.}{2012}]{Hobbs2012}
{Hobbs} G. {\rm et~al.}, 2012, \mnras, 427, 2780

\bibitem[\protect\citeauthoryear{{Hobbs}, {Lorimer}, {Lyne} \& {Kramer}}{{Hobbs} et~al.}{2005}]{Hobbs2005}
{Hobbs} G.,  {Lorimer} D.~R.,  {Lyne} A.~G.,    {Kramer} M.,  2005, \mnras, 360, 974

\bibitem[\protect\citeauthoryear{{Hobbs}, {Edwards} \& {Manchester}}{{Hobbs} et~al.}{2006}]{Hobbs2006}
{Hobbs} G.~B.,  {Edwards} R.~T.,    {Manchester} R.~N.,  2006, \mnras, 369, 655

\bibitem[\protect\citeauthoryear{{Holtsmark}}{{Holtsmark}}{1919}]{Holtsmark1919}
{Holtsmark} J.,  1919, Annalen der Physik, 363, 577

\bibitem[\protect\citeauthoryear{{Hotan}, {van Straten} \& {Manchester}}{{Hotan} et~al.}{2004}]{Hotan2004}
{Hotan} A.~W.,  {van Straten} W.,    {Manchester} R.~N.,  2004, \pasa, 21, 302

\bibitem[\protect\citeauthoryear{{Johnson} et~al.,}{{Johnson} et~al.}{2013}]{FermiB1821}
{Johnson} T.~J.,  et~al., 2013, ApJ, submitted (Broadband Pulsations from PSR~B1821$-$24)

\bibitem[\protect\citeauthoryear{{Johnston} \& {Bailes}}{{Johnston} \& {Bailes}}{1991}]{Johnston1991}
{Johnston} S.,  {Bailes} M.,  1991, \mnras, 252, 277


\bibitem[\protect\citeauthoryear{{Kehl} \& {Krieger}}{{Kehl} \& {Krieger}}{2012}]{KK2012}
{Kehl} M., {Krieger} A., 2012, Auswirkungen der Verletzung des Starken \"Aquivalenzprinzips auf die Dynamik von Bin\"arpulsaren im Gravitationsfeld der Milchstra{\ss}e und in Kugelsternhaufen. Bachelor thesis, University of Bonn, Germany, 2012. 

\bibitem[\protect\citeauthoryear{{Keith}, {Jameson}, {van Straten}, {Bailes}, {Johnston}, {Kramer}, {Possenti}, {Bates}, {Bhat}, {Burgay}, {Burke-Spolaor},
  {D'Amico}, {Levin}, {McMahon}, {Milia} \& {Stappers}}{{Keith}
  et~al.}{2010}]{HTRU1}
{Keith} M.~J. {\rm et~al.},  2010, \mnras, 409, 619

\bibitem[\protect\citeauthoryear{{Keith}, {Johnston}, {Bailes}, {Bates}, {Bhat}, {Burgay}, {Burke-Spolaor}, {D'Amico}, {Jameson}, {Kramer}, {Levin},
  {Milia}, {Possenti}, {Stappers}, {van Straten} \& {Parent}}{{Keith} et~al.}{2012}]{HTRU4}
{Keith} M.~J. {\rm et~al.},  2012, \mnras, 419, 1752

\bibitem[\protect\citeauthoryear{{Keith} et~al.,}{{Keith} et~al.}{2013}]{Keith2013}
{Keith} M.~J.,  et~al., 2013, \mnras, 429, 2161

\bibitem[\protect\citeauthoryear{{Kerr}}{{Kerr}}{2011}]{Kerr2011}
{Kerr} M.,  2011, \apj, 732, 38

\bibitem[\protect\citeauthoryear{{Kopeikin}}{{Kopeikin}}{1996}]{Kopeikin1996}
{Kopeikin} S.~M.,  1996, \apjl, 467, L93

\bibitem[\protect\citeauthoryear{{Lange}, {Camilo}, {Wex}, {Kramer}, {Backer}, {Lyne} \& {Doroshenko}}{{Lange} et~al.}{2001}]{Lange2001}
{Lange} C.,  {Camilo} F.,  {Wex} N.,  {Kramer} M.,  {Backer} D.~C.,  {Lyne} A.~G.,    {Doroshenko} O.,  2001, \mnras, 326, 274

\bibitem[\protect\citeauthoryear{{Lazaridis}, {Verbiest}, {Tauris}, {Stappers}, {Kramer}, {Wex}, {Jessner}, {Cognard}, {Desvignes}, {Janssen}, {Purver},
  {Theureau}, {Bassa} \& {Smits}}{{Lazaridis} et~al.}{2011}]{Lazaridis2011}
{Lazaridis} K. {\rm et~al.},  2011, \mnras, 414, 3134

\bibitem[\protect\citeauthoryear{{Levin} et~al.,}{{Levin} et~al.}{2013}]{Levin2013}
{Levin} L. {\rm et~al.}, 2013, \mnras, 434, 1387

\bibitem[\protect\citeauthoryear{{Lorimer} \& {Kramer}}{{Lorimer} \& {Kramer}}{2004}]{Handbook2004}
{Lorimer} D.~R.,  {Kramer} M.,  2004, {Handbook of Pulsar Astronomy}

\bibitem[\protect\citeauthoryear{{Lynch} et~al.,}{{Lynch} et~al.}{2013}]{Lynch2013}
{Lynch} R.~S.,  et~al., 2013, in IAU Symposium Vol.~291 of IAU Symposium, {The hunt for new pulsars with the Green Bank Telescope}.
pp 41--46

\bibitem[\protect\citeauthoryear{{Manchester} et~al.,}{{Manchester} et~al.}{2013}]{Manchester2013}
{Manchester} R.~N. {\rm et~al.}, 2013, \pasa, 30, 17

\bibitem[\protect\citeauthoryear{{Manchester}, {Hobbs}, {Teoh} \& {Hobbs}}{{Manchester} et~al.}{2005}]{PSRCAT}
{Manchester} R.~N.,  {Hobbs} G.~B.,  {Teoh} A.,    {Hobbs} M.,  2005, VizieR Online Data Catalog, 7245, 0

\bibitem[\protect\citeauthoryear{{Mihalas} \& {Binney}}{{Mihalas} \& {Binney}}{1981}]{Mihalas1981}
{Mihalas} D., {Binney} J., 1981, {Galactic astronomy: Structure and kinematics /2nd edition/}

\bibitem[\protect\citeauthoryear{{Ng} et~al.,}{{Ng} et~al.}{in prep}]{Ng}
{Ng} C. {\rm et~al.}, in prep

\bibitem[\protect\citeauthoryear{{Nolan} et~al.,}{{Nolan} et~al.}{2012}]{Fermi2FGL}
{Nolan} P.~L. {\rm et~al.}, 2012, \apjs, 199, 31

\bibitem[\protect\citeauthoryear{{Nomoto}, {Nariai} \& {Sugimoto}}{{Nomoto} et~al.}{1979}]{Nomoto1979}
{Nomoto} K.,  {Nariai} K.,    {Sugimoto} D.,  1979, \pasj, 31, 287

\bibitem[\protect\citeauthoryear{{Paczynski}}{{Paczynski}}{1990}]{Paczynski1990}
{Paczynski} B.,  1990, \apj, 348, 485

\bibitem[\protect\citeauthoryear{Peters}{Peters}{1964}]{Peters1964}
Peters P.~C.,  1964, Phys. Rev., 136, B1224

\bibitem[\protect\citeauthoryear{{Petroff}, {Keith}, {Johnston}, {van Straten} \& {Shannon}}{{Petroff} et~al.}{2013}]{Petroff2013}
{Petroff} E.,  {Keith} M.~J.,  {Johnston} S.,  {van Straten} W.,    {Shannon} R.~M.,  2013, \mnras

\bibitem[\protect\citeauthoryear{{Phinney}}{{Phinney}}{1992}]{Phinney1992}
{Phinney} E.~S.,  1992, Royal Society of London Philosophical Transactions Series A, 341, 39

\bibitem[\protect\citeauthoryear{{Pletsch} et~al.,}{{Pletsch} et~al.}{2012}]{Pletsch2012}
{Pletsch} H.~J. {\rm et~al.}, 2012, Science, 338, 1314

\bibitem[\protect\citeauthoryear{{Ray} et~al.,}{{Ray}  et~al.}{2011}]{Ray2011}
{Ray} P.~S. {\rm et~al.}, 2011, \apjs, 194, 17

\bibitem[\protect\citeauthoryear{{Ray}, {Ransom}, {Cheung}, {Giroletti}, {Cognard}, {Camilo}, {Bhattacharyya}, {Roy}, {Romani}, {Ferrara},
  {Guillemot}, {Johnston}, {Keith}, {Kerr}, {Kramer}, {Pletsch}, {Saz Parkinson} \& {Wood}}{{Ray} et~al.}{2013}]{Ray2013}
{Ray} P.~S. {\rm et~al.},  2013, \apjl, 763, L13

\bibitem[\protect\citeauthoryear{{Roberts}}{{Roberts}}{2013}]{Roberts2013}
{Roberts} M.~S.~E.,  2013, in IAU Symposium Vol.~291 of IAU Symposium, {Surrounded by spiders! New black widows and redbacks in the Galactic field}.
pp 127--132

\bibitem[\protect\citeauthoryear{{Romani}, {Filippenko}, {Silverman}, {Cenko}, {Greiner}, {Rau}, {Elliott} \& {Pletsch}}{{Romani} et~al.}{2012}]{Romani2012}
{Romani} R.~W.,  {Filippenko} A.~V.,  {Silverman} J.~M.,  {Cenko} S.~B., {Greiner} J.,  {Rau} A.,  {Elliott} J.,    {Pletsch} H.~J.,  2012, \apjl,
  760, L36

\bibitem[\protect\citeauthoryear{{Schnitzeler}}{{Schnitzeler}}{2012}]{Schnitzeler2012}
{Schnitzeler} D.~H.~F.~M.,  2012, \mnras, 427, 664

\bibitem[\protect\citeauthoryear{{Shklovskii}}{{Shklovskii}}{1970}]{Shk1970}
{Shklovskii} I.~S.,  1970, \sovast, 13, 562

\bibitem[\protect\citeauthoryear{{Spitkovsky}}{{Spitkovsky}}{2006}]{Spitkovsky2006}
{Spitkovsky} A.,  2006, \apjl, 648, L51

\bibitem[\protect\citeauthoryear{{Stairs}}{{Stairs}}{2003}]{Stairs2003}
{Stairs} I.~H.,  2003, Living Reviews in Relativity, 6, 5

\bibitem[\protect\citeauthoryear{{Stairs}, {Faulkner}, {Lyne}, {Kramer}, {Lorimer}, {McLaughlin}, {Manchester}, {Hobbs}, {Camilo}, {Possenti},
  {Burgay}, {D'Amico}, {Freire} \& {Gregory}}{{Stairs} et~al.}{2005}]{Stairs2005}
{Stairs} I.~H. {\rm et~al.},  2005, \apj, 632, 1060
 
\bibitem[\protect\citeauthoryear{{Staveley-Smith}, {Wilson}, {Bird}, {Disney}, {Ekers}, {Freeman}, {Haynes}, {Sinclair}, {Vaile}, {Webster} \&
  {Wright}}{{Staveley-Smith} et~al.}{1996}]{Multibeam1996}
{Staveley-Smith} L. {\rm et~al.},  1996, \pasa, 13, 243

\bibitem[\protect\citeauthoryear{{Taam} \& {van den Heuvel}}{{Taam} \& {van den Heuvel}}{1986}]{TvH1986}
{Taam} R.~E.,  {van den Heuvel} E.~P.~J.,  1986, \apj, 305, 235

\bibitem[\protect\citeauthoryear{{Tauris}}{{Tauris}}{2011}]{Tauris2011}
{Tauris} T.~M.,  2011, in {Schmidtobreick} L.,  {Schreiber} M.~R.,   {Tappert} C.,  eds, Evolution of Compact Binaries Vol.~447 of Astronomical Society of
  the Pacific Conference Series, {Five and a Half Roads to Form a Millisecond Pulsar}.
p.~285

\bibitem[\protect\citeauthoryear{{Tauris} \& {Bailes}}{{Tauris} \& {Bailes}}{1996}]{Tauris1996}
{Tauris} T.~M.,  {Bailes} M.,  1996, \aap, 315, 432

\bibitem[\protect\citeauthoryear{{Tauris}, {Langer} \& {Kramer}}{{Tauris} et~al.}{2012}]{Tauris2012}
{Tauris} T.~M.,  {Langer} N.,    {Kramer} M.,  2012, \mnras, 425, 1601

\bibitem[\protect\citeauthoryear{{Tauris} \& {Savonije}}{{Tauris} \& {Savonije}}{1999}]{Tauris1999}
{Tauris} T.~M.,  {Savonije} G.~J.,  1999, \aap, 350, 928

\bibitem[\protect\citeauthoryear{{Tauris} \& {van den Heuvel}}{{Tauris} \& {van den Heuvel}}{2006}]{Tauris2006}
{Tauris} T.~M.,  {van den Heuvel} E.~P.~J.,  2006, {Formation and evolution of compact stellar X-ray sources}.
pp 623--665

\bibitem[\protect\citeauthoryear{{Taylor}}{{Taylor}}{1992}]{Taylor1992}
{Taylor} J.~H.,  1992, Royal Society of London Philosophical Transactions Series A, 341, 117

\bibitem[\protect\citeauthoryear{{Taylor} \& {Cordes}}{{Taylor} \& {Cordes}}{1993}]{TC93}
{Taylor} J.~H.,  {Cordes} J.~M.,  1993, \apj, 411, 674 

\bibitem[\protect\citeauthoryear{{Thornton} et~al.,}{{Thornton} et~al.}{in prep}]{Thornton}
{Thornton} D. {\rm et~al.}, in prep

\bibitem[\protect\citeauthoryear{{Toscano}, {Sandhu}, {Bailes}, {Manchester}, {Britton}, {Kulkarni}, {Anderson} \& {Stappers}}{{Toscano}
  et~al.}{1999}]{Toscano1999}
{Toscano} M.,  {Sandhu} J.~S.,  {Bailes} M.,  {Manchester} R.~N.,  {Britton} M.~C.,  {Kulkarni} S.~R.,  {Anderson} S.~B.,    {Stappers} B.~W.,  1999,
  \mnras, 307, 925

\bibitem[\protect\citeauthoryear{{van Haaften}, {Nelemans}, {Voss} \& {Jonker}}{{van Haaften} et~al.}{2012}]{vHaaften2012}
{van Haaften} L.~M.,  {Nelemans} G.,  {Voss} R.,    {Jonker} P.~G.,  2012, \aap, 541, A22

\bibitem[\protect\citeauthoryear{{van Haaften}, {Nelemans}, {Voss}, {Wood} \& {Kuijpers}}{{van Haaften} et~al.}{2012}]{vanHaaften2012}
{van Haaften} L.~M.,  {Nelemans} G.,  {Voss} R.,  {Wood} M.~A.,    {Kuijpers} J.,  2012, \aap, 537, A104

\bibitem[\protect\citeauthoryear{{van Haasteren} et~al.,}{{van Haasteren} et~al.}{2011}]{vHaasteren2011}
{van Haasteren} R. {\rm et~al.}, 2011, \mnras, 414, 3117

\bibitem[\protect\citeauthoryear{{van Straten}}{{van Straten}}{2004}]{vStraten2004}
{van Straten} W.,  2004, \apjs, 152, 129

\bibitem[\protect\citeauthoryear{{Venter}, {Johnson} \& {Harding}}{{Venter} et~al.}{2012}]{Venter2012}
{Venter} C., 2012, \apj, 744, 34

\bibitem[\protect\citeauthoryear{{Verbiest}, {Lorimer} \& {McLaughlin}}{{Verbiest} et~al.}{2010}]{Verbiest2010}
{Verbiest} J.~P.~W.,  {Lorimer} D.~R.,    {McLaughlin} M.~A.,  2010, \mnras, 405, 564

\bibitem[\protect\citeauthoryear{{Weisberg} \& {Taylor}}{{Weisberg} \& {Taylor}}{1981}]{Weisberg1981}
{Weisberg} J.~M.,  {Taylor} J.~H.,  1981, General Relativity and Gravitation, 13, 1

\bibitem[\protect\citeauthoryear{{Yardley}, {Coles}, {Hobbs}, {Verbiest}, {Manchester}, {van Straten}, {Jenet}, {Bailes}, {Bhat}, {Burke-Spolaor},
  {Champion}, {Hotan}, {Oslowski}, {Reynolds} \& {Sarkissian}}{{Yardley} et~al.}{2011}]{Yardley2011}
{Yardley} D.~R.~B. {\rm et~al.},  2011, \mnras, 414, 1777

\bibitem[\protect\citeauthoryear{{You}, {Hobbs}, {Coles}, {Manchester}, {Edwards}, {Bailes}, {Sarkissian}, {Verbiest}, {van Straten}, {Hotan}, {Ord},
  {Jenet}, {Bhat} \& {Teoh}}{{You} et~al.}{2007}]{You2007}
{You} X.~P. {\rm et~al.},  2007, \mnras, 378, 493


\end{thebibliography}

\label{lastpage}

\end{document}